\documentclass{interact}

\usepackage{epstopdf}
\usepackage[caption=false]{subfig}

\usepackage[numbers,sort&compress]{natbib}
\bibpunct[, ]{[}{]}{,}{n}{,}{,}
\renewcommand\bibfont{\fontsize{10}{12}\selectfont}

\usepackage{cleveref}
\usepackage{hyperref}
\numberwithin{equation}{section}

\theoremstyle{plain}

\theoremstyle{definition}

\theoremstyle{remark}

\begin{document}

\articletype{}

\title{Algebraic discretization of time-independent Hamiltonian systems using a Lie-group/algebra approach}

\author{
\name{S. Bertrand\thanks{Email: sebastien.bertrand@fjfi.cvut.cz}}
\affil{Czech Technical University in Prague, Faculty of Nuclear Sciences and Physical Engineering, Department of Physics, B\v{r}ehov\'a 7, 115 19 Prague 1, Czech Republic}
}

\maketitle

\begin{abstract}
In this paper, time-independent Hamiltonian systems are investigated via a Lie-group/algebra formalism. The (unknown) solution linked with the Hamiltonian is considered to be a Lie-group transformation of the initial data, where the group parameter acts as the time. The time-evolution generator (i.e. the Lie algebra associated to the group transformation) is constructed at an algebraic level, hence avoiding discretization of the time-derivatives for the discrete case. This formalism makes it possible to investigate the continuous and discrete versions of time for time-independent Hamiltonian systems and no additional information on the system is required (besides the Hamiltonian itself and the initial conditions of the solution). When the time-independent Hamiltonian system is integrable in the sense of Liouville, one can use the action-angle coordinates to straighten the time-evolution generator and construct an exact scheme (i.e. a scheme without errors). In addition, a method to analyse the errors of approximative/numerical schemes is provided. These considerations are applied to well-known examples associated with the one-dimensional harmonic oscillator.
%
\end{abstract}


\section{Introduction}\label{SecIntro}
In physics, the concept of a discrete (space)time is gaining in interest \cite{Antippa07}. Theories, such as loop quantum gravity \cite{RV18}, use a discretization of the spacetime to solve some fundamental problems. In numerical analysis, the choice of discretization is critical to get good and efficient numerical approximations. The easiness with one gets approximations of solutions of very complex systems allows one to guess and explore behaviours of the systems even without being able to solve differential equations. However in general, from a physical or mathematical point of view, the breakdown from the continuous limit to the discrete case challenges conceptual considerations. The choice of the discretization scheme must be a smart one to get a consistent system and stable solutions, and to represent reality, especially. If one uses an arbitrary scheme, the numerical approximation could be completely wrong. Over the years, many approaches were taken to study time discretization of dynamic equations. Geometric approaches using symmetries seem to provide interesting results, see e.g. \cite{BV17,BJV19,HLW06} and references therein. 

Independently, the study of Lie groups and symmetries have been growing, especially in mathematical physics, see e.g. \cite{Newell85,Olver,RS02} and references therein. Among others, Lie symmetries have been successfully used  to derive dynamical equations (such as the Dirac equation \cite{Dirac48,Marchildon,Schwitchtenberg}), to reduce differential equations by facilitating the finding of their solutions (see e.g. \cite{Bertrand17,CW99,LRT20}) and even in the context of numerical analysis, e.g. \cite{BJV19,BV17,BCW06,BRW08,HLW06,KO04}. A particularly interesting theorem was provided by Noether \cite{Noether}, linking conservation laws to Lie symmetries. The conservation laws (or integrals of motion in the context of Hamiltonian systems) represent quantities that are preserved along any solution. These quantities led to the study of integrable systems in the sense of solitons (where a system of nonlinear partial differential equations possesses an infinite number of conservation laws \cite{Newell85}) and in the Liouville sense for Hamiltonian systems. In the case where the Hamiltonian is time-independent, integrability is defined by the existence of $N$ integrals of motion that are functionally independent and in involution with respect to the Poisson bracket, where $N$ is half the number of dimensions of the phase space. Integrable Hamiltonian systems possess a remarkable property, the existence of action-angle coordinates \cite{Arnold,Liouville}. Furthermore, some integrable Hamiltonian systems possess additional integrals of motion, they are called superintegrable. Maximally superintegrable systems possess the maximal number of integrals of motion, i.e. $2N-1$, and they possess the characteristic that bounded trajectories are closed and periodic \cite{BS19,MSW15,MPW13}.

The goal of this paper is to investigate an algebraic discretization of time for time-independent Hamiltonian systems. In order to do so, we propose a Lie-group/algebra approach such that time can be treated as a group parameter. We ensure that the Lie-group transformation and its Lie algebra satisfy many intrinsic conditions, such as the canonicality and the invariance of the integrals of motion under the group transformation. In the case where the time-independent system is integrable, we can rectify the evolution generator using the action-angle coordinates. This leads to a method to obtain exact schemes. In addition, we study non-exact schemes. We propose a method to obtain schemes based on the above formalism and we also investigate the errors of consistent schemes and provide methods to reduce the errors.

The paper is structured as follow. In section 2, we construct the formalism, in a first time for the continuous case, and then for the discrete case of time-independent Hamiltonian systems. In section 3, we investigate the special case of time-independent integrable Hamiltonian systems. For these considerations, we provide a step-by-step method to construct exact schemes and we show the method for the well-known case of the one-dimensional harmonic oscillator. In section 4, we apply the formalism of section 2 to learn more about non-exact (approximative) schemes. We illustrate these results using three different schemes for the one-dimensional harmonic oscillator: the exact scheme, the Euler method and the discrete gradient method. In section 5, we give some conclusions and provide some future perspective.

\section{Construction of the time-evolution generator}\label{SecForm}
Let us consider a time-independent Hamiltonian $H=H(\vec{q},\vec{p})$ defined on a $2N$-dimensional phase space $\mathcal{M}$, where the positions and the momenta are described respectively by the canonical variables $q_j$ and $p_j$, $j=1,...,N$. The associated equations of motion are given by
\begin{eqnarray}
\dot{q}_j=\frac{\partial H}{\partial p_j}=\lbrace q_j,H\rbrace,\qquad \dot{p}_j=-\frac{\partial H}{\partial q_j}=\lbrace p_j,H\rbrace,\label{eqsMot}
\end{eqnarray}
where the dot represents the total derivative with respect to time, i.e. $\dot{f}(t)=\frac{df(t)}{dt}$, and the bracket $\lbrace\cdot,\cdot\rbrace$ is the usual Poisson bracket, i.e.
\begin{eqnarray}
\lbrace a,b\rbrace=\sum_{j=1}^N\left(\frac{\partial a}{\partial q_j}\frac{\partial b}{\partial p_j}-\frac{\partial b}{\partial q_j}\frac{\partial a}{\partial p_j}\right),\label{PB}
\end{eqnarray}
The canonical variables $\vec{q}$, $\vec{p}$ satisfy the relations
\begin{eqnarray}
\lbrace q_i,p_j\rbrace=\delta_{ij},\qquad \lbrace q_i,q_j\rbrace=\lbrace p_i,p_j\rbrace=0,\qquad i,j=1,...,N, \label{PBrel}
\end{eqnarray}
where $\delta_{ij}$ is the Kronecker delta. One should note that since the Hamiltonian $H$ is time-independent, then the system possesses at least one integral of motion, namely the Hamiltonian $H$ itself.

In addition, we will assume that the solution the Hamiltonian system (\ref{eqsMot}) with the initial condition
\begin{eqnarray}
q_j(t_0)=c_j,\qquad p_j(t_0)=c_{j+N},\qquad j=1,...,N,
\end{eqnarray}
exists, is unique, smooth and equivalent to a curve in a $2N+1$ dimension space (the phase space augmented with the time dimension). 

\subsection{The continuous case}
Let us assume the existence of a Lie-group transformation $g_\Delta$ allowing an advance in time $\Delta$. We require that the Lie-group transformation is linear in time, that is successive iterations are equivalent to one iteration of the sum of their steps, $g_{\Delta_1}\circ g_{\Delta_2}=g_{\Delta_1+\Delta_2}$. However, instead of using an explicit advance in time, i.e. $g_\Delta\circ f(t)=f(t+\Delta)$, we will consider an implicit version, that is the transformation will solely depend on the parameter $\Delta$ and the current-time variables $\vec{q}(t)$ and $\vec{p}(t)$, i.e.
\begin{eqnarray}
g_\Delta\circ q_j\equiv Q_j=F_j(\vec{q},\vec{p};\Delta),\qquad g_\Delta\circ p_j\equiv P_j=G_j(\vec{q},\vec{p};\Delta),\label{Gtrans}
\end{eqnarray}
where the variables $Q_j$ and $P_j$ are the advanced-time variables and the transformation applies on the current-time variables $\vec{q}(t)$ and $\vec{p}(t)$. The identity transformation is given by taking the limit of $\Delta$ to $0$,
\begin{eqnarray}
\lim_{\Delta\rightarrow0}\left(F_j(\vec{q},\vec{p};\Delta)\right)=q_j,\qquad \lim_{\Delta\rightarrow0}\left(G_j(\vec{q},\vec{p};\Delta)\right)=p_j,\label{Idlim}
\end{eqnarray}
and the inverse transformation by setting $\Delta$ to $-\Delta$, i.e. $g_\Delta^{-1}=g_{-\Delta}$.

The associated infinitesimal deformation, the Lie algebra, is the vector field
\begin{eqnarray}
\mathfrak{g}=\sum_{j=1}^N\left(\xi_j(\vec{q},\vec{p})\frac{\partial}{\partial q_j}+\eta_j(\vec{q},\vec{p})\frac{\partial}{\partial p_j}\right),\label{Gvec}
\end{eqnarray}
where the functions $\xi_j$ and $\eta_j$ are usually defined by the relations
\begin{eqnarray}
\xi_j(\vec{q},\vec{p})=\lim_{\Delta\rightarrow0}\left(\frac{\partial}{\partial\Delta}F_j(\vec{q},\vec{p};\Delta)\right),\qquad \eta_j(\vec{q},\vec{p})=\lim_{\Delta\rightarrow0}\left(\frac{\partial}{\partial\Delta}G_j(\vec{q},\vec{p};\Delta)\right).\label{Vlim}
\end{eqnarray}
The group transformation (\ref{Gtrans}) can be written using the step $\Delta$ and the evolution generator as
\begin{eqnarray}
g_\Delta = \exp(\Delta \mathfrak{g}).
\end{eqnarray}

The group transformation (\ref{Gtrans}) and its Lie algebra (\ref{Gvec}) must satisfy many conditions. One of them is that they must leave the integrals of motion $I_k=I_k(\vec{q},\vec{p})$ invariant. This consequence can be expressed as
\begin{eqnarray}
\mathfrak{g}\circ I_k=0\qquad \mbox{mod}\quad I_k=0,\label{SymCon}
\end{eqnarray}
which is the condition for the Lie algebra $\mathfrak{g}$ to be a symmetry of the algebraic set of integrals of motion. One can look e.g. in \cite{Olver} for details on symmetries of sets of algebraic equations.

In addition, transformation (\ref{Gtrans}) must be canonical. We require that the same relations as in (\ref{PBrel}) must be satisfied regardless of the advance in time, that is
\begin{eqnarray}
\lbrace Q_i,P_j\rbrace=\delta_{ij},\qquad\lbrace Q_i,Q_j\rbrace=\lbrace P_i,P_j\rbrace=0,\qquad i,j=1,...,N.
\end{eqnarray}
For the mixed bracket $\lbrace Q_i,P_j\rbrace$, we have
\begin{eqnarray}
\delta_{ij}&=&\lbrace Q_i,P_j\rbrace=\lbrace F_i(\vec{q},\vec{p};\Delta),G_j(\vec{q},\vec{p};\Delta)\rbrace\nonumber\\
&=&\lbrace\exp(\Delta \mathfrak{g})\circ q_i,\exp(\Delta \mathfrak{g})\circ p_j\rbrace.\label{PBcons}
\end{eqnarray}
However, working in the Lie-group formalism, with the exponentials, is much harder than working in the Lie-algebra formalism, so we want to decompose the exponential and want to know what are the implications of (\ref{PBcons}) on the vector field $\mathfrak{g}$. From one side, the composition of two canonical transformations remains canonical. From the other side, the group transformation can be seen as applying $n$ times the infinitesimal deformation $1+\frac{\Delta}{n}\mathfrak{g}$ and taking the limit of $n$ to infinity. Hence, if the infinitesimal deformation is canonical, so is the group transformation. This implies that the conditions from equation (\ref{PBcons}) are equivalent to
\begin{eqnarray}
\delta_{ij}&=&\left\lbrace\left(1+\frac{\Delta}{n}\mathfrak{g}\right)\circ q_i,\left(1+\frac{\Delta}{n}\mathfrak{g}\right)\circ p_j\right\rbrace\nonumber\\
&=&\left\lbrace q_i+\frac{\Delta}{n}\xi_i,p_j+\frac{\Delta}{n}\eta_j\right\rbrace\nonumber\\
&=&\lbrace q_i,p_j\rbrace+\frac{\Delta}{n}\left(\lbrace q_i,\eta_j\rbrace+\lbrace\xi_i,p_j\rbrace\right)+O^2\left(\frac{1}{n}\right)\nonumber\\
&=&\delta_{ij}+\frac{\Delta}{n}\left(\frac{\partial\eta_j}{\partial p_i}+\frac{\partial\xi_i}{\partial q_j}\right)+O^2\left(\frac{1}{n}\right).\nonumber
\end{eqnarray}
Hence we have the conditions
\begin{eqnarray}
\frac{\partial\eta_j}{\partial p_i}+\frac{\partial\xi_i}{\partial q_j}=0\label{Vcan1}
\end{eqnarray}
and similarly for the relations $\lbrace Q_i,Q_j\rbrace$ and $\lbrace P_i,P_j\rbrace$:
\begin{eqnarray}
\frac{\partial \xi_j}{\partial p_i}-\frac{\partial \xi_i}{\partial p_j}=0,\qquad \frac{\partial \eta_j}{\partial q_i}-\frac{\partial \eta_i}{\partial q_j}=0.\label{Vcan2}
\end{eqnarray}
The general solution to the canonical conditions (\ref{Vcan1}) and (\ref{Vcan2}) is
\begin{eqnarray}
\xi_j(\vec{q},\vec{p})=\frac{\partial h(\vec{q},\vec{p})}{\partial p_j},\qquad \eta_j(\vec{q},\vec{p})=-\frac{\partial h(\vec{q},\vec{p})}{\partial q_j},\label{CanCon}
\end{eqnarray}
where $h(\vec{q},\vec{p})$ is an arbitrary function of the phase-space coordinates.

Combining the symmetry condition (\ref{SymCon}) with the canonical condition (\ref{CanCon}),  we see that the function $h$ must Poisson-bracket commute with all the integrals of motion, i.e.
\begin{eqnarray}
\mathfrak{g}\circ I_k=\sum_{j=0}^N\left(\frac{\partial h}{\partial p_j}\frac{\partial I_k}{\partial q_j}-\frac{\partial h}{\partial q_j}\frac{\partial I_k}{p_j}\right)=\lbrace I_k,h\rbrace=0\qquad\mbox{mod}\quad I_k=0.
\end{eqnarray}
In 1 dimension, it is trivial to see that the only arbitrary function that can commute with the Hamiltonian $H$ is a function of the Hamiltonian. Hence, the vector field takes the form
\begin{eqnarray}
\mathfrak{g}&=&h'(H)\left(\frac{\partial H}{\partial p_1}\frac{\partial}{\partial q_1}-\frac{\partial H}{\partial q_1}\frac{\partial}{\partial p_1}\right)=h'(H)\left(\dot{q}_1\frac{\partial}{\partial p_1}+\dot{p}_1\frac{\partial}{\partial p_1}\right)\nonumber\\
&=&h'(H)\frac{d}{dt},\nonumber
\end{eqnarray}
which is equivalent to the symmetry of translation in time multiplied by a constant $h'(H)$. In fact, by setting  $h=H$, we can recover the time evolution of the system, the equations of motion, by going back to the group transformation, i.e.
\begin{eqnarray}
\frac{dq_1}{\xi_1(q_1,p_2)}=d\Delta=\frac{dp_1}{\eta_1(q_1,p_1)}\nonumber
\end{eqnarray}
or more explicitly
\begin{eqnarray}
\frac{dq_1}{d\Delta}=\frac{\partial H}{\partial p_1}=\xi(q_1,p_1),\qquad \frac{dp}{d\Delta}=-\frac{\partial H}{\partial q_1}=\eta(q_1,p_1),\qquad \frac{dq_1}{dp_1}=-\frac{\frac{\partial H}{\partial p_1}}{\frac{\partial H}{\partial q_1}},
\end{eqnarray}
where the group parameter $\Delta$ acts as the time. Thus, the infinitesimal generator of evolution in time takes the form
\begin{eqnarray}
\mathfrak{g}=\frac{\partial H}{\partial p_1}\frac{\partial}{\partial q_1}-\frac{\partial H}{\partial q_1}\frac{\partial}{\partial p_1},\qquad\mbox{such that}\quad \mathfrak{g}\circ\phi=\lbrace\phi,H\rbrace.
\end{eqnarray}

In higher dimension, the argument is less straightforward. For maximally superintegrable systems, only the Hamiltonian can Poisson-bracket commute with all the integrals of motion, hence the derivation is similar, but with more variables to consider. For non-maximally superintegrable systems, some integrals of motion may Poisson-bracket with all other integrals, however, only the Hamiltonian will ensure commutation, recover the time-evolution in the form of the equations of motion and represent a time-translation symmetry. Therefore, the infinitesimal generator of the group transformation (\ref{Gtrans}) is given uniquely by
\begin{eqnarray}
\mathfrak{g}=\sum_{j=1}^N\left(\frac{\partial H}{\partial p_j}\frac{\partial}{\partial q_j}-\frac{\partial H}{\partial q_j}\frac{\partial}{\partial p_j}\right),\qquad \mathfrak{g}\circ\phi=\lbrace\phi,H\rbrace
\end{eqnarray}
and will be called the (time-)evolution generator. It represents the flow of the Hamiltonian vector field and is already know in the literature, see e.g. \cite{Olver}. Here, we approached it without using total-derivatives in time, which will be useful in the next section to avoid the definition of discrete time-derivative.

\subsection{The discrete case}
A very similar derivation can be done in the discrete case, but problems arise when one wants to take the limit of $\Delta$ to $0$. Hence, we need to use a different approach in the discrete case when the limit $\Delta\rightarrow0$ is considered. In the continuous version, we used the limit at two places, that is in equations (\ref{Idlim}) and (\ref{Vlim}). For the equations (\ref{Idlim}), the identity of the Lie-group transformation must exist. By taking the composition of an advance in time and a return in time, i.e.
\begin{eqnarray}
g_\Delta^{-1}\circ g_\Delta=\exp(-\Delta \mathfrak{g})\circ\exp(\Delta \mathfrak{g})=\exp((\Delta-\Delta)\mathfrak{g})=I,
\end{eqnarray}
we can see that the identity exists and is well-defined without using the limit. For equations (\ref{Vlim}), we use the alternative definitions:
\begin{eqnarray}
\xi_j(\vec{q},\vec{p})=\exp(-\Delta \mathfrak{g})\circ\frac{\partial}{\partial\Delta}F_j(\vec{q},\vec{p};\Delta),\label{DisXi}\\
\eta_j(\vec{q},\vec{p})=\exp(-\Delta \mathfrak{g})\circ\frac{\partial}{\partial\Delta}G_j(\vec{q},\vec{p};\Delta),\label{DisEta}
\end{eqnarray}
which are equivalent in the continuous case, but not in the discrete case. It can be seen as applying the group transformation, then differentiating with respect to $\Delta$ and applying the inverse transformation. If one uses the limit $\Delta\rightarrow0$ instead of the above definition, then any numerical/approximation scheme that is consistent would satisfy the constraints. We shall discuss it further when we consider non-exact schemes.

Thus, if one follows the same steps as previously, but with the relations (\ref{DisXi}) and (\ref{DisEta}) for the vector fields, one will get that the evolution generator takes the form
\begin{eqnarray}
\mathfrak{g}=\sum_{j=1}^N\left(\frac{\partial H}{\partial p_j}\frac{\partial}{\partial q_j}-\frac{\partial H}{\partial q_j}\frac{\partial}{\partial p_j}\right),\qquad \mathfrak{g}\circ\phi=\lbrace\phi,H\rbrace,
\end{eqnarray}
which is the same as in the continuous case. Hence, the exact scheme of evolution of the system can be obtained by returning to the group transformation. However, we saw that getting back to the group transformation is equivalent to solving the equations of motion, which is not necessarily an easy task. This approach does not help to solve the equations of motion, but it provides with a guide for the evolution of the system. 

In the discrete case, interpolations between the points are not physically interesting. The continuous trajectory generated by the Lie group is to be considered a tool, not 
a real behaviour. This tool allows us to have an arbitrary mesh, which may be useful to jump over singularities.

Overall, the group transformation $g_\Delta$ can be seen as a transformation of the initial conditions at $t=t_0$ to new``initial conditions'' at $t=t_0+\Delta$. By applying repetitively the group transformation, we can map the solution or a subset of it.
%

\section{Exact schemes for integrable systems}\label{SecExact}
Let us consider that the system (\ref{eqsMot}) is integrable, i.e. there exist at least $N$ integrals of motion $I_k$ that are functionally independent and in involution with respect to the Poisson bracket (\ref{PB}). According to the Liouville--Arnold theorem \cite{Arnold,Liouville}, integrable systems possess a distinguished set of coordinates called the action-angle coordinates. The action-angle coordinates can be obtained via a canonical transformation where the associated equations of motion take the form
\begin{eqnarray}
\frac{dz_k}{dt}=\nu_k(\vec{I}),\qquad \frac{dI_k}{dt}=0,\qquad k=1,...,N.\label{AAeqm}
\end{eqnarray}
Even if the $\nu_k$ are functions of $I_k$, they can be treated as constants since the $I_k$ are constant over time. Moreover, the Liouville-Arnold theorem states that the trajectories in the phase space are diffeomorphic to an $N$-dimensional torus under some additional conditions. However, we will ignore this geometric interpretation and weaken the constrains on the system and its solutions. Hence, by the action-angle coordinates, we solely refer to the system of coordinates leading to the equations of motion (\ref{AAeqm}).

In the action-angle coordinates, the evolution generator is straightened to
\begin{eqnarray}
\mathfrak{g}=\sum_{k=1}^N\nu_k(\vec{I})\frac{\partial}{\partial z_k},
\end{eqnarray}
for which the associated group transformation is
\begin{eqnarray}
z_k\rightarrow z_k+\nu_k\Delta,\qquad I_k\rightarrow I_k.
\end{eqnarray}

The action-angle coordinates are a particularly good choice for discretization. This comes from the fact that $N$ integrals of motion will be preserved in the discretization by definition. Furthermore, since the frequencies $\nu_k$ are treated as constants, there is no arbitrarity in the choice of the discretization scheme, conversely with the choice made in the traditional methods, such as Runge--Kutta methods. Hence, if one knows the algebraic transformations between the original phase-space coordinates and the action-angle coordinates, then one can go to the action-angle coordinates to discretize and come back to the original system to obtain an exact scheme.

\subsection{Method using a generating function of type 2}\label{SSecMethAA}
We assume that $N$ integrals of motion in involution are known explicitly. If the integrals of motion are not known, there exist many ways to find them. There is a tremendous number of articles in the literature trying the classify all integrable systems, see e.g. \cite{BS19, MSW15, MPW13} and references therein. One could also try to find them by brute force or using physical intuition. Symbolic softwares, such as Maple \cite{Maple}, are able to find in some cases Lie point symmetries of a system of differential equations. By combinating it with Noether's theorem \cite{Noether} (making a link between symmetries and conservation laws), one can find integrals of motion.

By performing the following steps, one should be able (at least in theory) to obtain the time-discretized version of the equations of motion associated with the Hamiltonian system. Here, we use a generating function to get the relations between the action and the angle variables, but other methods exist. In addition, one should note that the action-angle coordinates are not unique. One can always use a different function of the same integral and obtain different coordinates, which will lead to equivalent results.

\bigskip
\noindent\textbf{Steps:}
\begin{enumerate}
\item Assign the commuting integrals of motion $I_k$ as the action coordinates, i.e.
\begin{eqnarray}
I_k=I_k(\vec{q},\vec{p}).\label{eqsStep1}
\end{eqnarray}
\item From equations (\ref{eqsStep1}), solve the momenta $p_k$ in terms of the generalized coordinates $q_k$ and the action coordinates $I_k$, i.e.
\begin{eqnarray}
p_k=s_k(\vec{q},\vec{I}).
\end{eqnarray}
\item Find a generating function $K$ of second type satisfying the system of partial differential equations
\begin{eqnarray}
\frac{\partial K(\vec{q},\vec{I})}{\partial q_k}=s_k(\vec{q},\vec{I}).
\end{eqnarray}
\item By substituting the solution of the generating function $K$ into the equations
\begin{eqnarray}
z_k=\theta_k(\vec{q},\vec{I})=\frac{\partial K(\vec{q},\vec{I})}{\partial I_k},
\end{eqnarray}
get the relations for the angle coordinates $z_k$.
\item Solve the inverse transformation of $\theta_k$ to get the $q_k$ in terms of the action-angle coordinates, i.e.
\begin{eqnarray}
q_k=r_k(\vec{z},\vec{I}).
\end{eqnarray}
\item Calculate the frequencies $\nu_k$ using the relations of the angle coordinates $z_k$ and the original equations of motion, i.e.
\begin{eqnarray}
\nu_k=\left.\sum_{j=1}^N\frac{\partial \theta_k(\vec{q},\vec{I})}{\partial q_j}\frac{\partial H(\vec{q},\vec{p})}{\partial p_j}\right\vert_{t=t_0}
\end{eqnarray}
\item Calculate the constant values of the integrals of motion, i.e.
\begin{eqnarray}
\gamma_k=\left.I_k(\vec{q},\vec{p})^{\phantom{^2}}\right\vert_{t=t_0}
\end{eqnarray}
\end{enumerate}
\textbf{Results:}
The advance-time original coordinates $Q_k,P_k$ can be expressed using the current-time original coordinates $\vec{q},\vec{p}$ as
\begin{eqnarray}
Q_k=r_k\left(\theta_1(\vec{q},\vec{\gamma})+\nu_1\Delta,...,\theta_N(\vec{q},\vec{\gamma})+\nu_N\Delta,\gamma_1,...,\gamma_N\right),\label{ExactX}\\
P_k=s_k\left(\theta_1(\vec{q},\vec{\gamma})+\nu_1\Delta,...,\theta_N(\vec{q},\vec{\gamma})+\nu_N\Delta,\gamma_1,...,\gamma_N\right),\label{ExactP}
\end{eqnarray}
where $\Delta$ is the time step. The results remain true regardless of the method used to find the action-angle coordinates.

\subsection{Example --- The 1-dimensional harmonic oscillator}\label{SSecExAA}
Let us consider the 1-dimensional harmonic oscillator given by the Hamiltonian
\begin{eqnarray}
H=\frac{\rho^2}{2m}+\frac{k}{2}\zeta^2,
\end{eqnarray}
where $\rho(\tau)$ is the momentum and $\zeta(\tau)$ is the position in time $\tau$. The equations of motion are given by
\begin{eqnarray}
\dot{\zeta}=\frac{\rho}{m},\qquad \dot{\rho}=-k\zeta.
\end{eqnarray}
In order to simplify the notation, we will absorb the mass $m$ and the spring constant $k$ using the (extended) canonical transformation \cite{Antippa07}
\begin{eqnarray}
\zeta(\tau)=\frac{1}{\sqrt{k}}\;x(t),\quad \rho(\tau)=\sqrt{m}\;p(t),\quad \tau=\sqrt{\frac{m}{k}}\;t,\nonumber
\end{eqnarray}
such that the new Hamiltonian becomes
\begin{eqnarray}
H=\frac{p^2+x^2}{2}\label{Hex}
\end{eqnarray}
together with the new equations of motion
\begin{eqnarray}
\frac{dx}{dt}=\frac{\partial H}{\partial p}=p,\qquad \frac{dp}{dt}=-\frac{\partial H}{\partial x}=-x.
\end{eqnarray}
The analytic solution of this system is well-known and given by a rotation transformation in the phase space
\begin{eqnarray}
x(t)&=&x_0\cos\left(t\right)+p_0\sin\left(t\right),\label{solx}\\
p(t)&=&p_0\cos\left(t\right)-x_0\sin\left(t\right),\label{solp}
\end{eqnarray}
where $x_0$ and $p_0$ represent the initial position and momentum.

\bigskip
\noindent\textbf{Step 1:}\\
We set $H(x,p)=E$ to be the integral of motion used, i.e. the action coordinate.

\bigskip
\noindent\textbf{Step 2:}\\
We can solve $p$ in terms of $x$ and $E$, i.e.
\begin{eqnarray}
p=\epsilon\sqrt{2E-x^2},\qquad \epsilon^2=1.
\end{eqnarray}

\bigskip
\noindent\textbf{Step 3:}\\
We can obtain the generating function by integration,
\begin{eqnarray}
K(x,E)&=&\int p dx=\int \epsilon\sqrt{2E-x^2} dx\nonumber\\
&=&\frac{\epsilon}{2}x\sqrt{2E-x^2}+\epsilon E\arctan\left(\frac{x}{\sqrt{2E-x^2}}\right).\nonumber
\end{eqnarray}

\bigskip
\noindent\textbf{Step 4:}\\
The angle coordinate $\theta$ is given by
\begin{eqnarray}
\theta&=&\frac{\partial K}{\partial E}=\epsilon \arctan\left(\frac{x}{\sqrt{2E-x^2}}\right)\nonumber\\
&=&\arctan\left(\frac{x}{p}\right),\label{thetaof}
\end{eqnarray}
together with the action variable
\begin{eqnarray}
E=\frac{p^2+x^2}{2}.\label{rof}
\end{eqnarray}

\bigskip
\noindent\textbf{Step 5:}\\
From equations (\ref{thetaof}) and (\ref{rof}), we can see that this transformation is linked with the polar coordinates, i.e. by inverting the transformation, we get
\begin{eqnarray}
x=\sqrt{2E}\sin(\theta),\qquad p=\sqrt{2E}\cos(\theta).\label{xietaof}
\end{eqnarray}

\bigskip
\noindent\textbf{Step 6:}\\
To obtain the frequency of the system, one can check the equations of motion in the action-angle coordinates, which take the form
\begin{eqnarray}
\frac{dE}{dt}=0,\qquad \frac{d\theta}{dt}=1.\label{AAeqsmot}
\end{eqnarray}

\bigskip
\noindent\textbf{Step 7:}\\
The integral of motion is a constant given by 
\begin{eqnarray}
E=\frac{p_0^2+x_0^2}{2}
\end{eqnarray}
at $t=0$.

\bigskip
\noindent\textbf{Result:}\\
Therefore, after some algebraic manipulations, we get that the exact scheme using (\ref{ExactX}-\ref{ExactP}) is given by
\begin{eqnarray}
X=x_0\cos(\Delta)+p_0\sin(\Delta),\qquad P=p_0\cos(\Delta)-x_0\sin(\Delta).\label{Exact}
\end{eqnarray}
In addition, the discrete equations of motion take the form
\begin{eqnarray}
\frac{X-x}{\Delta}=\frac{x_0(\cos(\Delta)-1)+p_0\sin(\Delta)}{\Delta},\\
\frac{P-p}{\Delta}=\frac{p_0(\cos(\Delta)-1)-x_0\sin(\Delta)}{\Delta}.
\end{eqnarray}
By taking the limit $\Delta\rightarrow0$, we get back the equations of motion of the continuous case, which makes the scheme consistent. The discretization scheme of the one-dimensional harmonic oscillator corroborates the results found in \cite{Cieslinski09} using a different approach.

It is also possible the to apply repetitively the transformations (\ref{Exact}) with a constant step $\Delta$, such that it is possible to get an equivalent of integration of the equations motion, i.e. after applying the transformation $n$-times (where $n$ is an arbitrary integer), we get after some algebraic transformations
\begin{eqnarray}
x(n\Delta)=x_0\cos(n\Delta)+p_0\sin(n\Delta),\qquad p(n\Delta)=p_0\cos(n\Delta)-x_0\sin(n\Delta),
\end{eqnarray}
where $n\Delta$ can be seen as the time $t$, i.e.
\begin{eqnarray}
t=\lim_{\stackrel{n\rightarrow\infty}{\Delta\rightarrow0}}n\Delta
\end{eqnarray}
Similar results can be obtained using inhomogeneous step sizes.

\section{Applications for non-exact schemes}\label{NonExact}
Let us consider an approximative scheme
\begin{eqnarray}
Q_j=F_j(\vec{q},\vec{p};\Delta),\qquad P_j=G_j(\vec{q},\vec{p};\Delta),
\end{eqnarray}
which is not necessarily a group transformation with the parameter $\Delta$. We can write this transformation as
\begin{eqnarray}
\psi_\Delta=g_\Delta+w_\Delta,
\end{eqnarray}
such that the approximated advanced values  $Q_j$ and $P_j$ are
\begin{eqnarray}
Q_j=\psi_\Delta\circ q_j,\qquad P_j=\psi_\Delta\circ p_j,
\end{eqnarray}
where $g_\Delta$ is the exact evolution transformation, i.e.
\begin{eqnarray}
g_\Delta=\exp(\Delta \mathfrak{g}),\qquad \mathfrak{g}=\sum_{j=1}^N\left(\frac{\partial H}{\partial p_j}\frac{\partial}{\partial q_j}-\frac{\partial H}{\partial q_j}\frac{\partial}{\partial p_j}\right),
\end{eqnarray}
and $w_\Delta$ is the local error of the step taking the form of a vector field using the current-time variables,
\begin{eqnarray}
w_\Delta=\sum_{j=1}^N\alpha_j(\vec{q},\vec{p};\Delta)\frac{\partial}{\partial q_j}+\beta(\vec{q},\vec{p};\Delta)\frac{\partial}{\partial p_j}.
\end{eqnarray}
If the transformation $\psi_\Delta$ is consistent, then the error $w_\Delta$ can be expressed using $\Delta^2$-terms and higher-order terms in $\Delta$, that is the Taylor expansion takes the form
\begin{eqnarray}
w_\Delta=\sum_{k=2}^\infty\frac{\Delta^k}{k!}v_k,\qquad v_k=\left.\frac{\partial^k w_\Delta}{\partial\Delta^k}\right\vert_{\Delta=0},\qquad v_0=v_1=0.
\end{eqnarray}

In this formalism, we know $\psi_\Delta$ and $\mathfrak{g}$, but a priori not $g_\Delta=\exp(\Delta \mathfrak{g})$, $w_\Delta$ and $v_k$. (One should note that $g_\Delta$ is known using $\mathfrak{g}$, but it may be hard to compute explicitly.) The lower-order $v_k$ can be computed from the known quantity
\begin{eqnarray}
\psi_{-\Delta}\circ\frac{\partial}{\partial\Delta}\psi_\Delta &=& \left(\exp(-\Delta \mathfrak{g})+w_{-\Delta}\right)\circ\left( \mathfrak{g}\circ\exp(\Delta \mathfrak{g})+\frac{\partial w_\Delta}{\partial \Delta}\right)\nonumber\\
&=&\mathfrak{g}+\exp(-\Delta \mathfrak{g})\circ\frac{\partial w_\Delta}{\partial\Delta}+w_{-\Delta}\circ \mathfrak{g}\circ\exp(\Delta \mathfrak{g})+w_{-\Delta}\circ \frac{\partial w_\Delta}{\partial\Delta}.\label{Err}
\end{eqnarray}
By expanding in $\Delta$ on both sides, we get for each coefficient of $\Delta$ the following relations:
\begin{eqnarray}
&\Delta^1:&\,\left.\frac{\partial}{\partial\Delta}\left(\psi_{-\Delta}\circ\frac{\partial}{\partial\Delta}\psi_\Delta\right)\right\vert_{\Delta=0}=v_2,\nonumber\\
&\Delta^2:&\,\left.\frac{\partial^2}{\partial\Delta^2}\left(\psi_{-\Delta}\circ\frac{\partial}{\partial\Delta}\psi_\Delta\right)\right\vert_{\Delta=0}=v_3-2\mathfrak{g}\circ v_2+v_2\circ \mathfrak{g},\nonumber\\
&\Delta^3:&\,\left.\frac{\partial^3}{\partial\Delta^3}\left(\psi_{-\Delta}\circ\frac{\partial}{\partial\Delta}\psi_\Delta\right)\right\vert_{\Delta=0}=v_4-3\mathfrak{g}\circ v_3+v_3\circ \mathfrak{g}+3v_2^2+3\mathfrak{g}^2\circ v_2+3v_2\circ \mathfrak{g}^2\nonumber\\
&\vdots&\nonumber
\end{eqnarray}
From the $\Delta^1$--equation, we can obtain the second-order error $v_2$. Then, we can compute the third-order error $v_3$ from $v_2$ and the $\Delta^2$--equation. Then, $v_4$ from $\Delta^3$ and so on. Therefore, the error of any scheme can be predicted analytically.

Being able to know what are the errors on a scheme allows to control with a better precision the numerical trajectories without even knowing the solution. Here, we propose two options to improve the numerical results:
\begin{itemize}
\item \textbf{Find the associated invariants of the error}\\
An invariant $\phi$ of the leading-order error $w_k$ will be more precise, that is because
\begin{eqnarray}
w_k\circ\phi=0,
\end{eqnarray}
hence the error will be diminished. However, some information would be lost since the number of functionally independent invariant is always lower than the number of variables.
\item \textbf{Increase the order of the error}\\
By subtracting the lowest-order error(s) $w_k$ to the transformation $\phi$, you can raise the order of the error, and get a more precise scheme. However, for geometric schemes, to increase the order of the error may break some qualitative properties.
\end{itemize}
A problem arises from the fact that the deformations of variables are only tangent to the solution (unless the variables are the action-angle coordinates). Since the deformations are tangent to the solution, the lower-order errors tend to spread into the higher-order terms and it gets harder to interpret correctly the errors. However, for the leading-order error, this method seems accurate, at least for non-stiff systems. As proposed in the conclusions, a better way to interpret geometrically the errors would help in this matter.


One should note that in general, the Euler method is equivalent to approximating the Lie-group transformation to a linear operator, that is
\begin{eqnarray}
Q_j=\exp(\Delta \mathfrak{g})\circ q_j\approx (1+\Delta \mathfrak{g})\circ q_j=q_j+\Delta\frac{\partial H}{\partial p_j}\\
\Rightarrow \frac{Q_j-q_j}{\Delta}\approx\frac{\partial H(\vec{q},\vec{p})}{\partial p_j}
\end{eqnarray}
and similarly for $P_j$. One can truncate the exponential at a higher order to generate schemes, however, the results will depend on higher partial derivatives of the Hamiltonian. The Runge--Kutta method can be obtained by further expanding the partial derivatives using Taylor series \cite{BFB16,Kutta01,Runge95}.

\subsection{Examples of schemes of the 1-dimensional harmonic oscillator}
Throughout the rest of this section, we will solely focus on examples of schemes for the 1-dimensional harmonic oscillator with the variables rescaled to absorb to the parameters of the system, i.e.
\begin{eqnarray}
H=\frac{p^2+x^2}{2}.\label{Hex2}
\end{eqnarray}
The evolution generator of this Hamiltonian is given by
\begin{eqnarray}
\mathfrak{g}=p\frac{\partial}{\partial x}-x\frac{\partial}{\partial p}.
\end{eqnarray}
To illustrate the numerical results, we consider the trajectory starting at $x(0)=1$ and $p(0)=0$ until it reaches $t=20$. The continuous version is represented by a continuous red line and the discrete version is represented by blue dots, using iteration $\Delta=0.1$ in the software Maple \cite{Maple}. 

\subsubsection{The exact scheme}
As a first example, we will consider the exact scheme from section \ref{SecExact}. The iteration scheme is given by
\begin{eqnarray}
X=x\cos(\Delta)+p\sin(\Delta),\qquad P=p\cos(\Delta)-x\sin(\Delta),
\end{eqnarray}
which suggests that the discrete equations of motion take the form
\begin{eqnarray}
\frac{X-x}{\Delta}=\frac{x\cos(\Delta)+p\sin(\Delta)-x}{\Delta},\qquad \frac{P-p}{\Delta}=\frac{p\cos(\Delta)-x\sin(\Delta)-p}{\Delta}.
\end{eqnarray}
By direct substitution into equation (\ref{Err}), we obtain that the error $w_\Delta$ is null, as expected.

\begin{figure}[h!]
\centering
\caption{}
\includegraphics[scale=0.4]{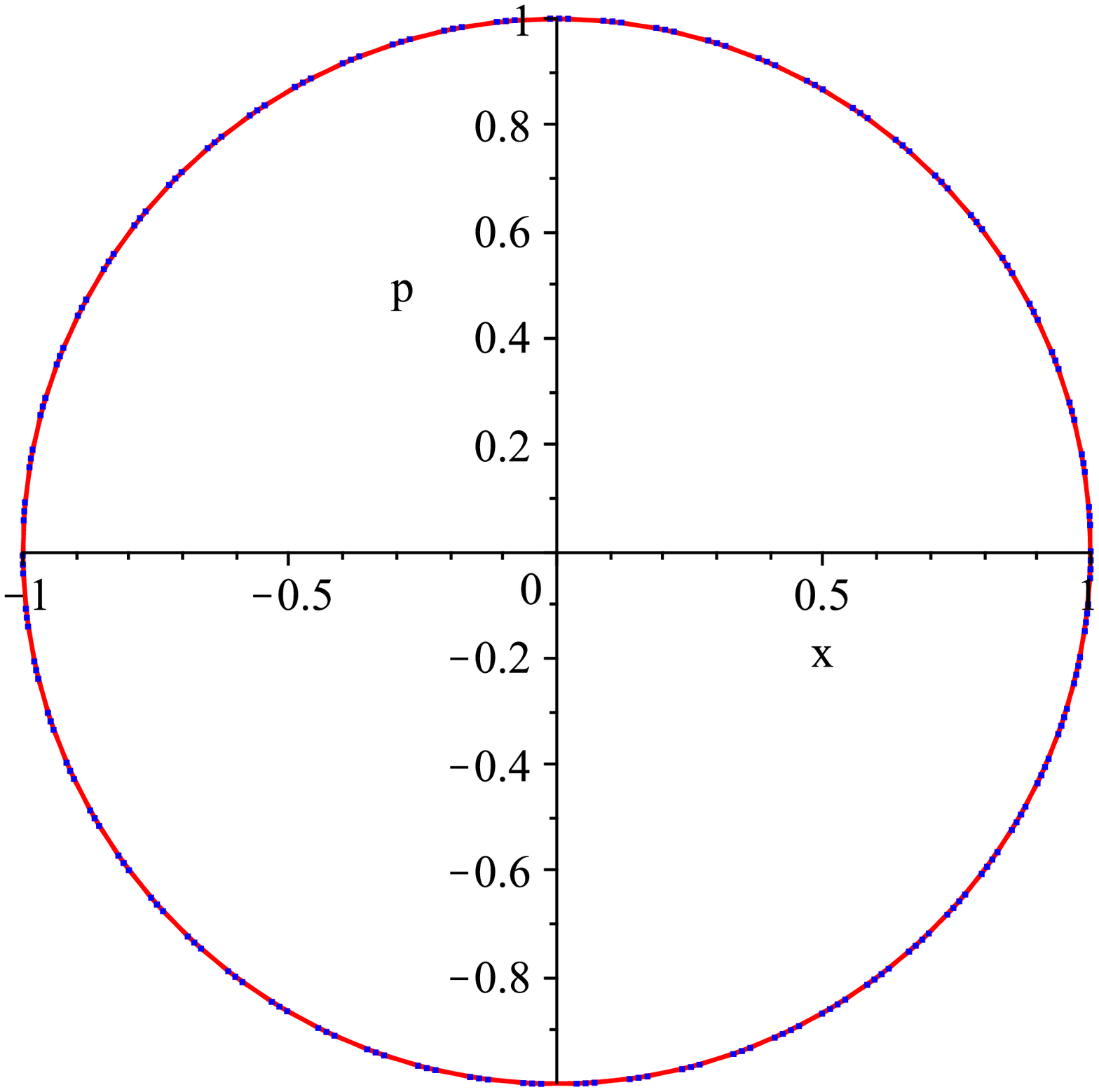}
\label{Exact-phase}
\end{figure}

\begin{figure}[h!]
\centering
\caption{}
\includegraphics[scale=0.7]{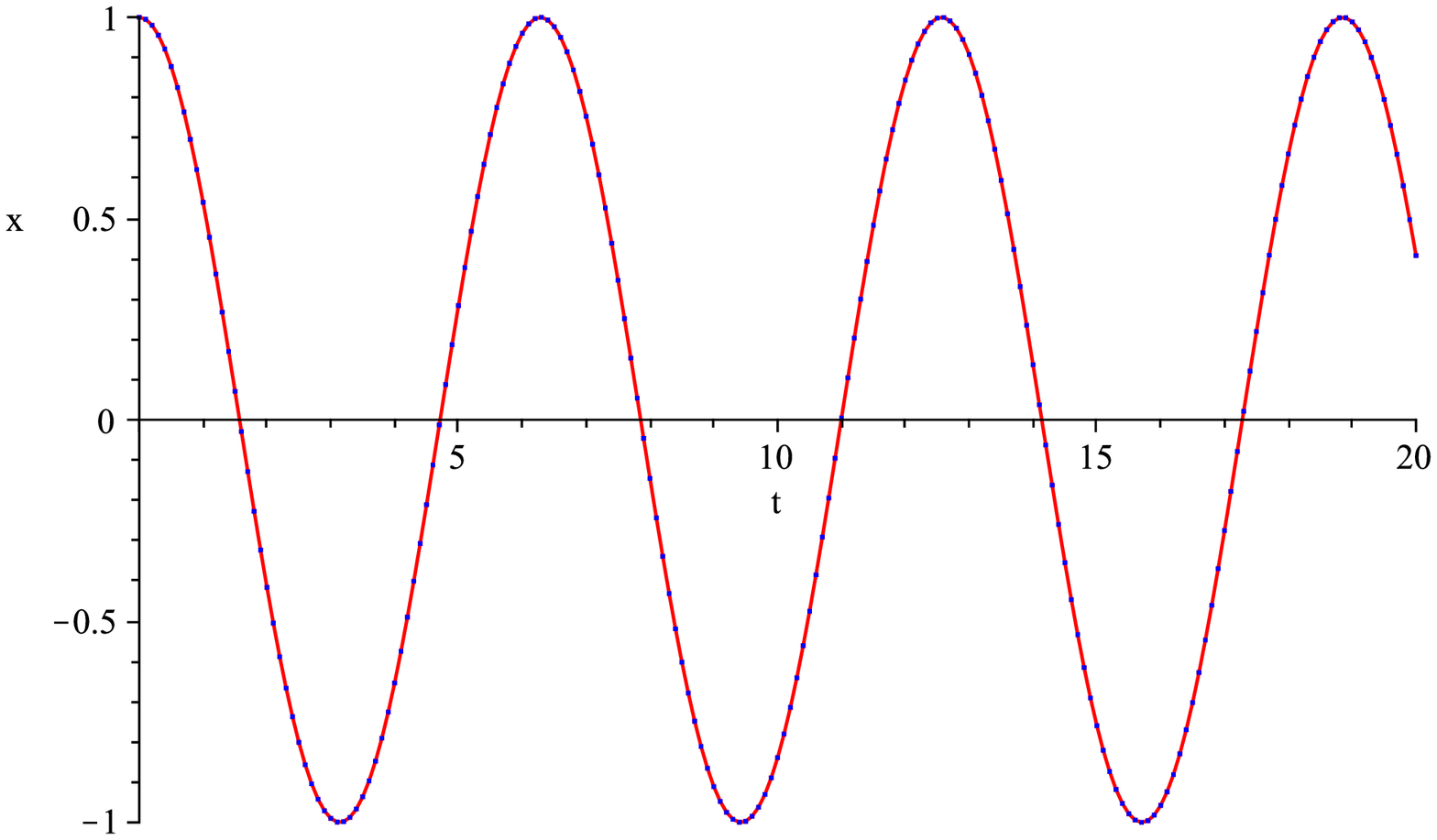}
\label{Exact-x}
\end{figure}

\begin{figure}[h!]
\centering
\caption{}
\includegraphics[scale=0.7]{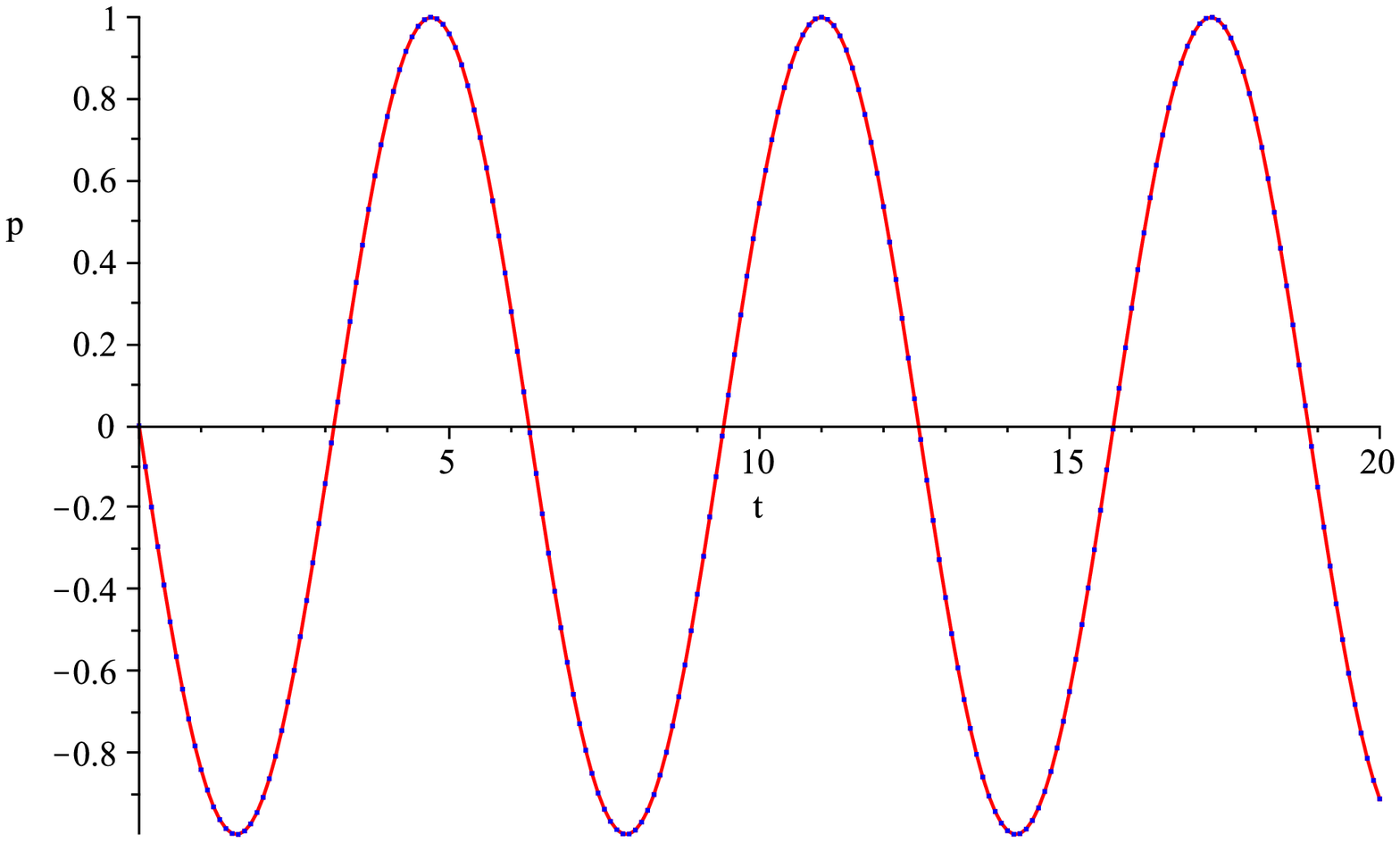}
\label{Exact-p}
\end{figure}

\begin{figure}[h!]
\centering
\caption{}
\includegraphics[scale=0.7]{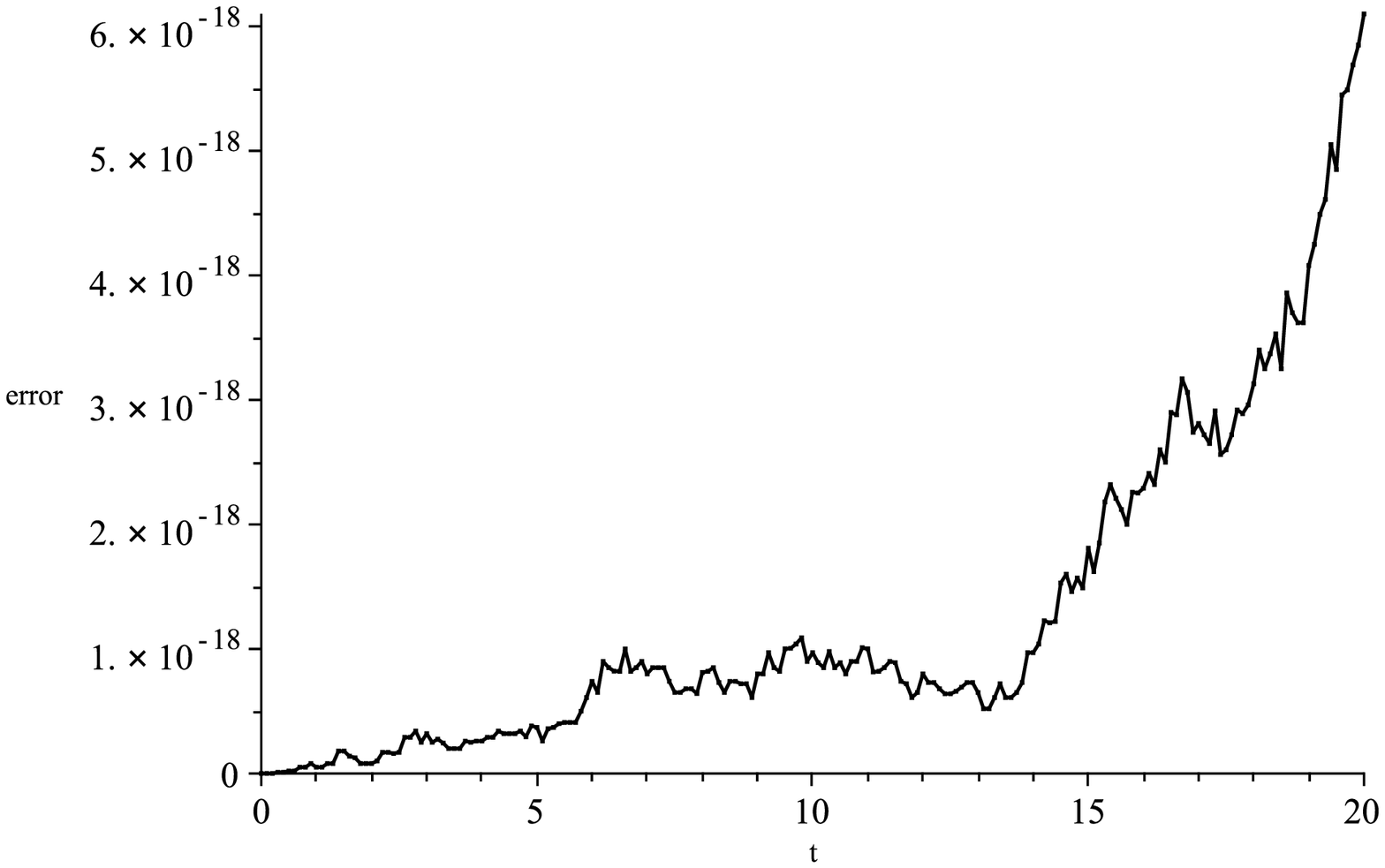}
\label{Exact-error}
\end{figure}

Figure \ref{Exact-phase} represents the trajectory in the phase space, Figure \ref{Exact-x} represents the evolution of $x$ in time and Figure \ref{Exact-p} represents the evolution of $p$ in time. Figure \ref{Exact-error} shows the error over time in the phase space,
\begin{eqnarray}
\sigma(x,p)=(x(t)-x(n\Delta))^2+(p(t)-p(n\Delta))^2,
\end{eqnarray}
such that $t=n\Delta$. In this scheme, the errors come from the approximation of the software.


\pagebreak

~

\pagebreak

\subsubsection{The Euler method}
The discrete equations of motion for the Euler method are
\begin{eqnarray}
\frac{X-x}{\Delta}=p,\qquad \frac{P-p}{\Delta}=-x
\end{eqnarray}
and the iteration scheme is given by
\begin{eqnarray}
X=x+\Delta p,\qquad P=p-\Delta x,
\end{eqnarray}
which is equivalent to the truncation of the exponential $\exp(\Delta \mathfrak{g})$ after the linear term in $\Delta$.

By direct substitution into equation (\ref{Err}), we obtain 
\begin{eqnarray}
\psi_{-\Delta}\circ\frac{\partial\psi_\Delta}{\partial\Delta}=(p+\Delta x)\frac{\partial}{\partial x}+(\Delta p-x)\frac{\partial}{\partial p}.
\end{eqnarray}
We can calculate the errors, that is
\begin{eqnarray}
v_0&=&v_1=0,\nonumber\\
v_2&=&x\frac{\partial}{\partial x}+p\frac{\partial}{\partial p},\nonumber\\
v_3&=&p\frac{\partial}{\partial x}+x\frac{\partial}{\partial p},\nonumber\\
v_4&=&-x\frac{\partial}{\partial x}-p\frac{\partial}{\partial p},\nonumber\\
&\vdots&\nonumber
\end{eqnarray}
The second-order error $v_2$ represents a scaling transformation, i.e.
\begin{eqnarray}
x\rightarrow e^\epsilon x\qquad p\rightarrow e^\epsilon p.\label{EulerTrans}
\end{eqnarray}
Hence, any function of the ratio $\frac{x}{p}$ will be an invariant of the error deformation $v_2$. 

\begin{figure}[h!]
\centering
\caption{}
\includegraphics[scale=0.4]{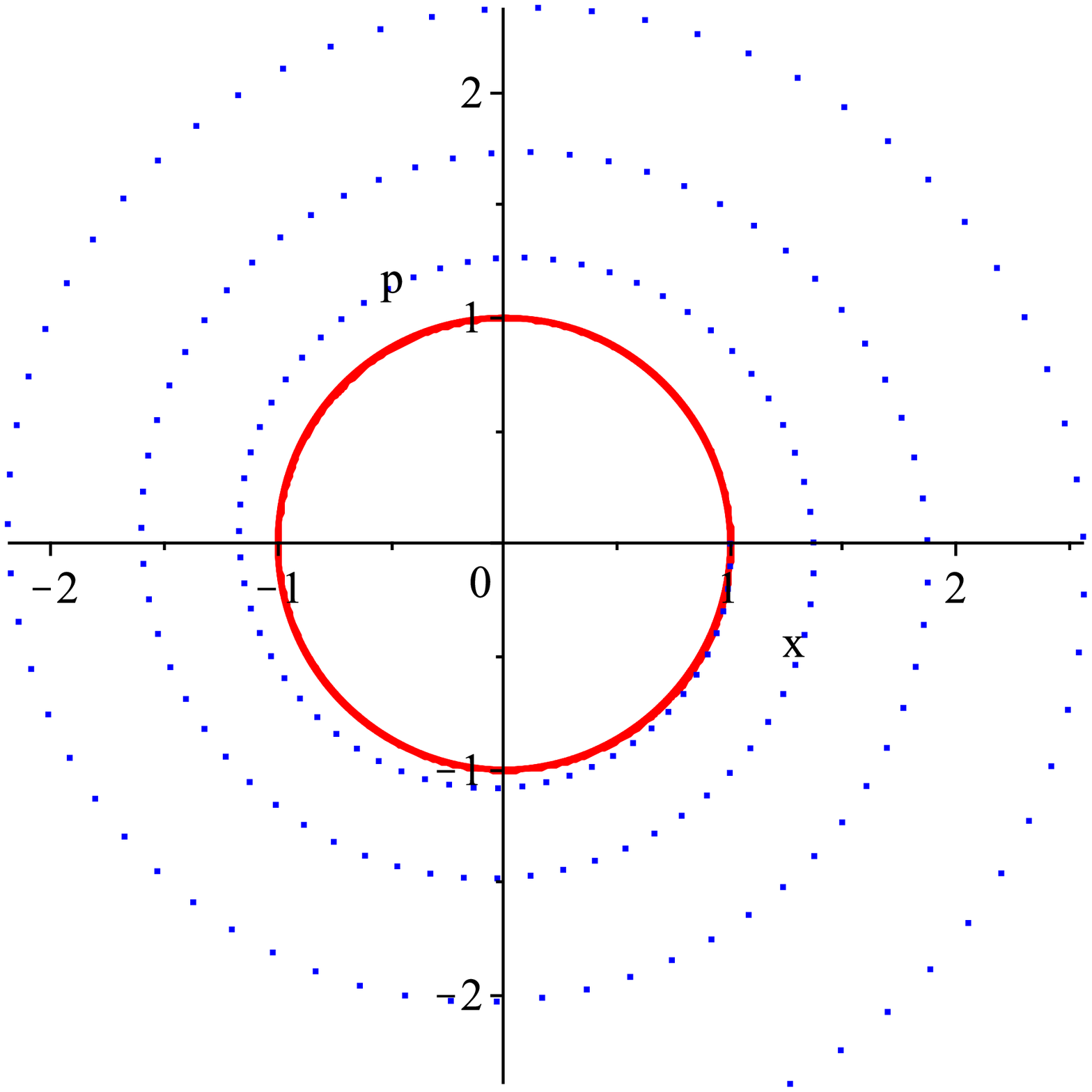}
\label{Euler-phase}
\end{figure}

\begin{figure}[h!]
\centering
\caption{}
\includegraphics[scale=0.7]{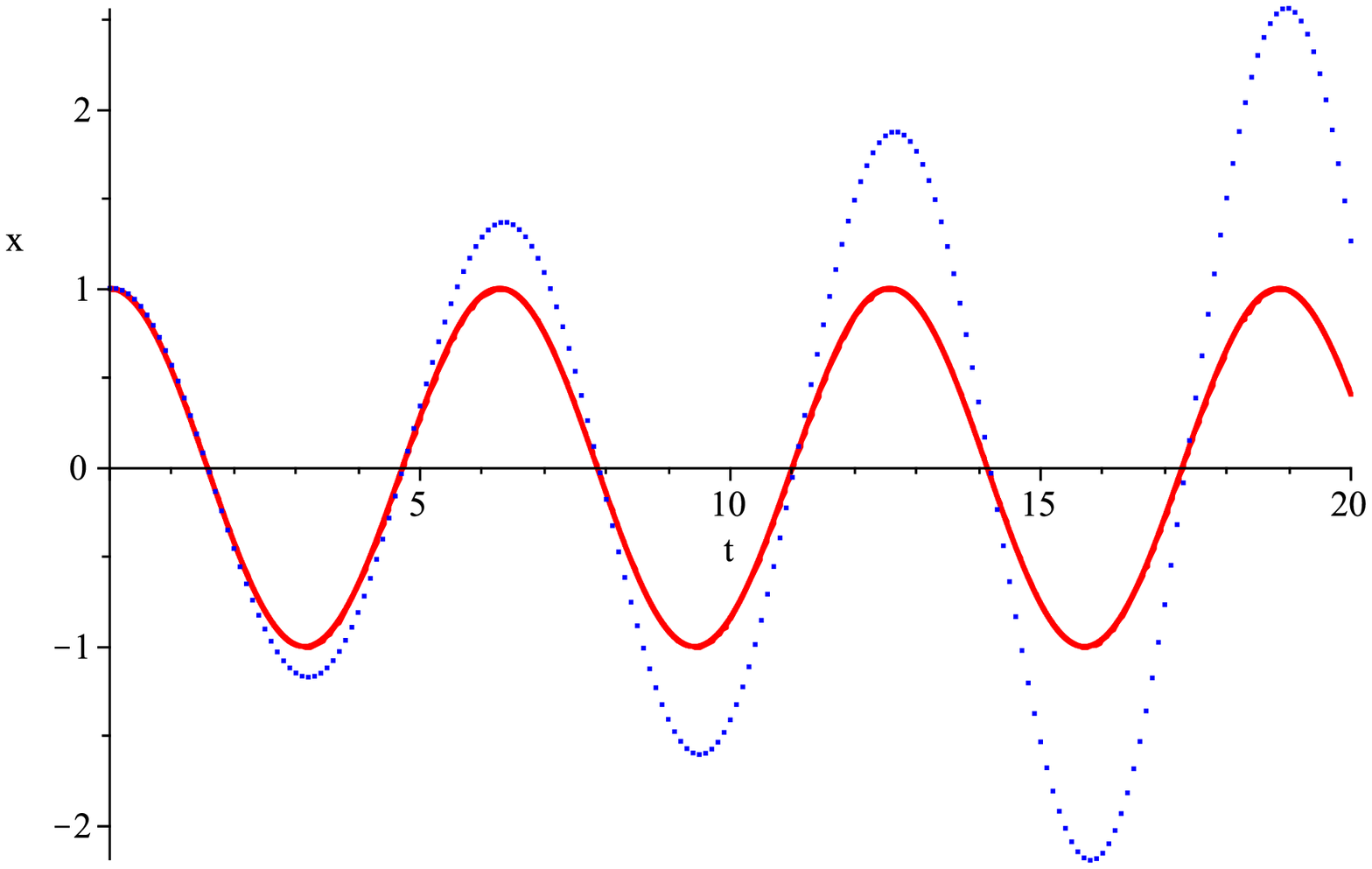}
\label{Euler-x}
\end{figure}

\begin{figure}[h!]
\centering
\caption{}
\includegraphics[scale=0.7]{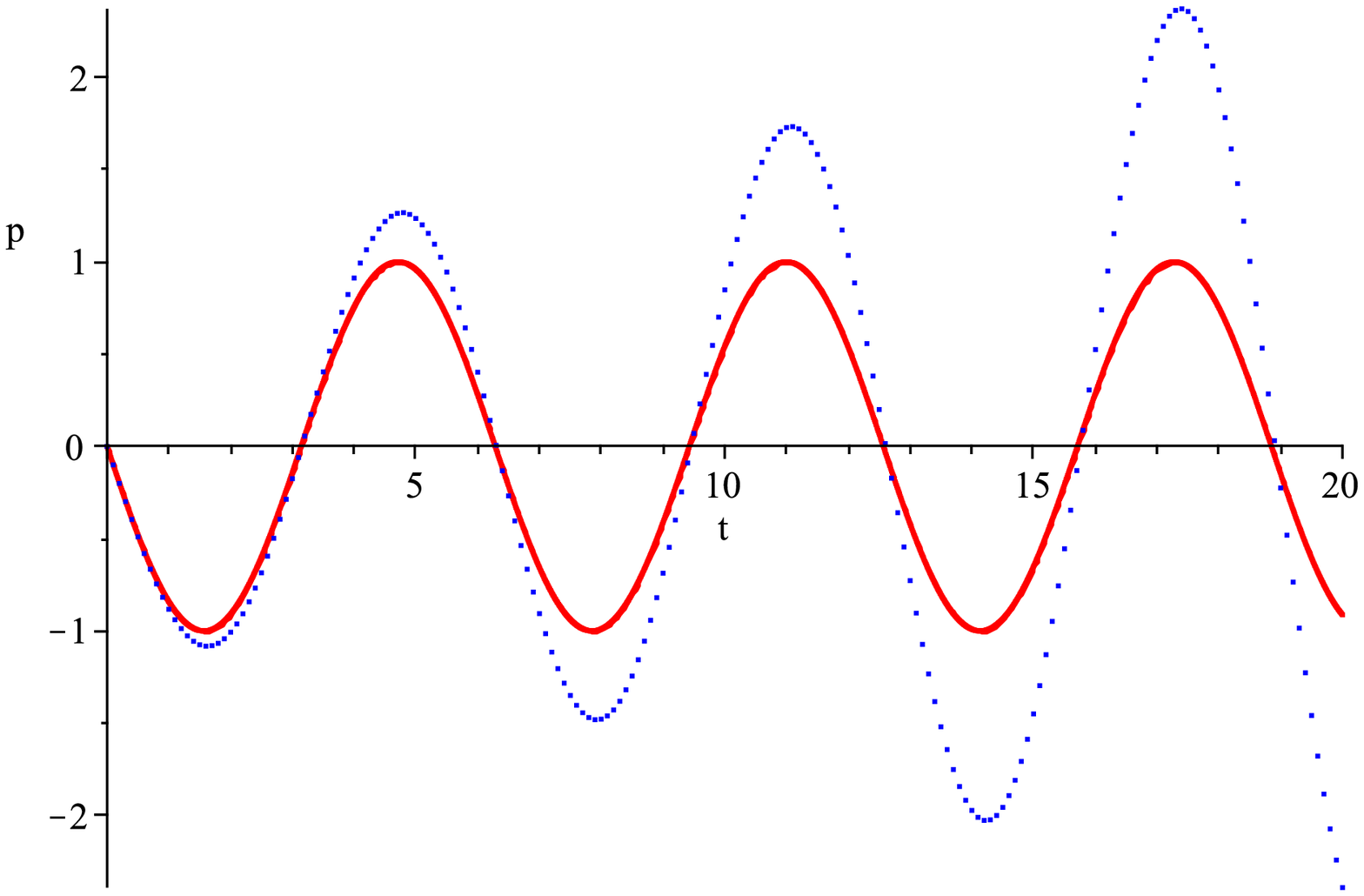}
\label{Euler-p}
\end{figure}

\begin{figure}[h!]
\centering
\caption{}
\includegraphics[scale=0.7]{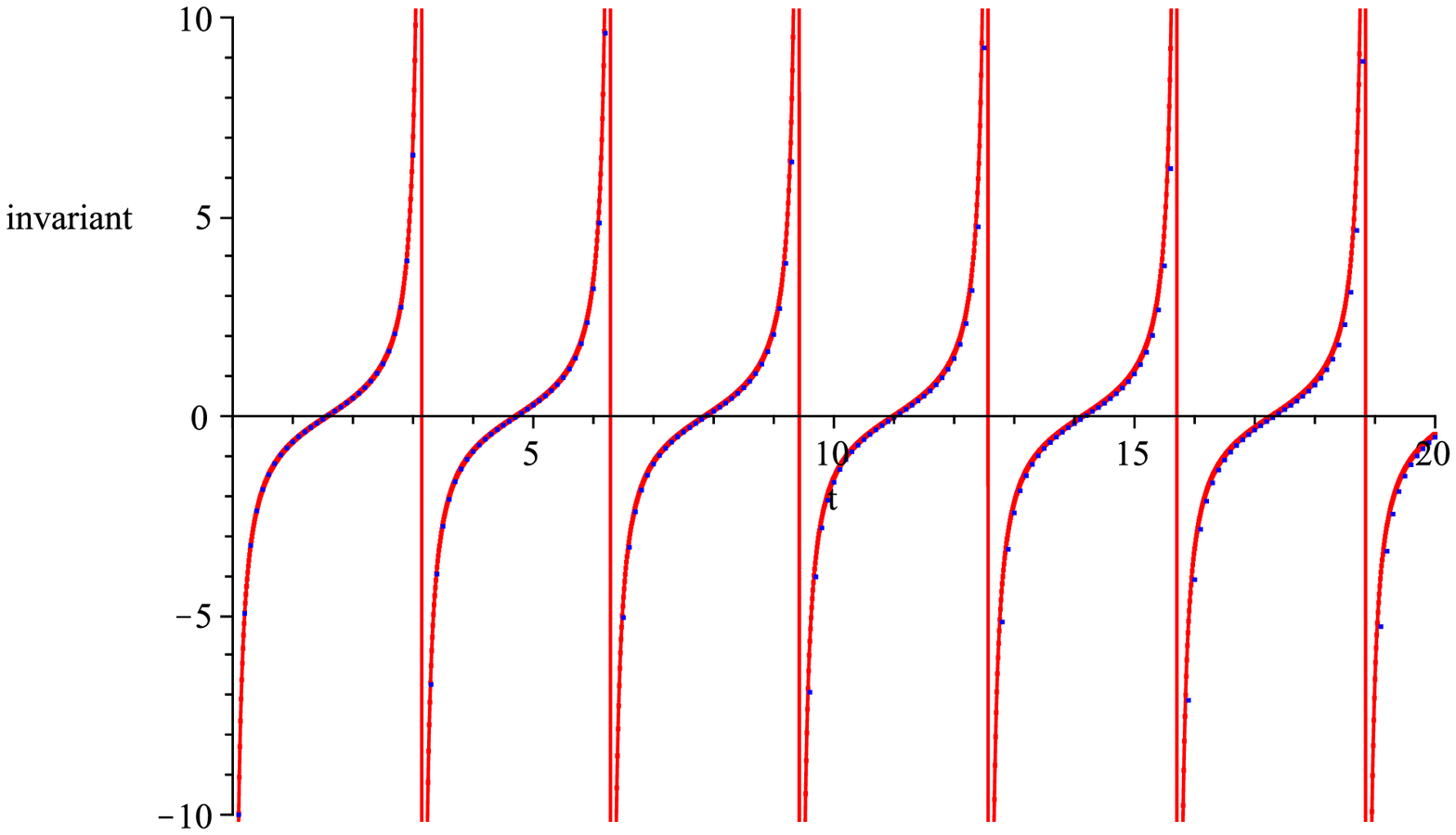}
\label{Euler-inv}
\end{figure}

\begin{figure}[h!]
\centering
\caption{}
\includegraphics[scale=0.7]{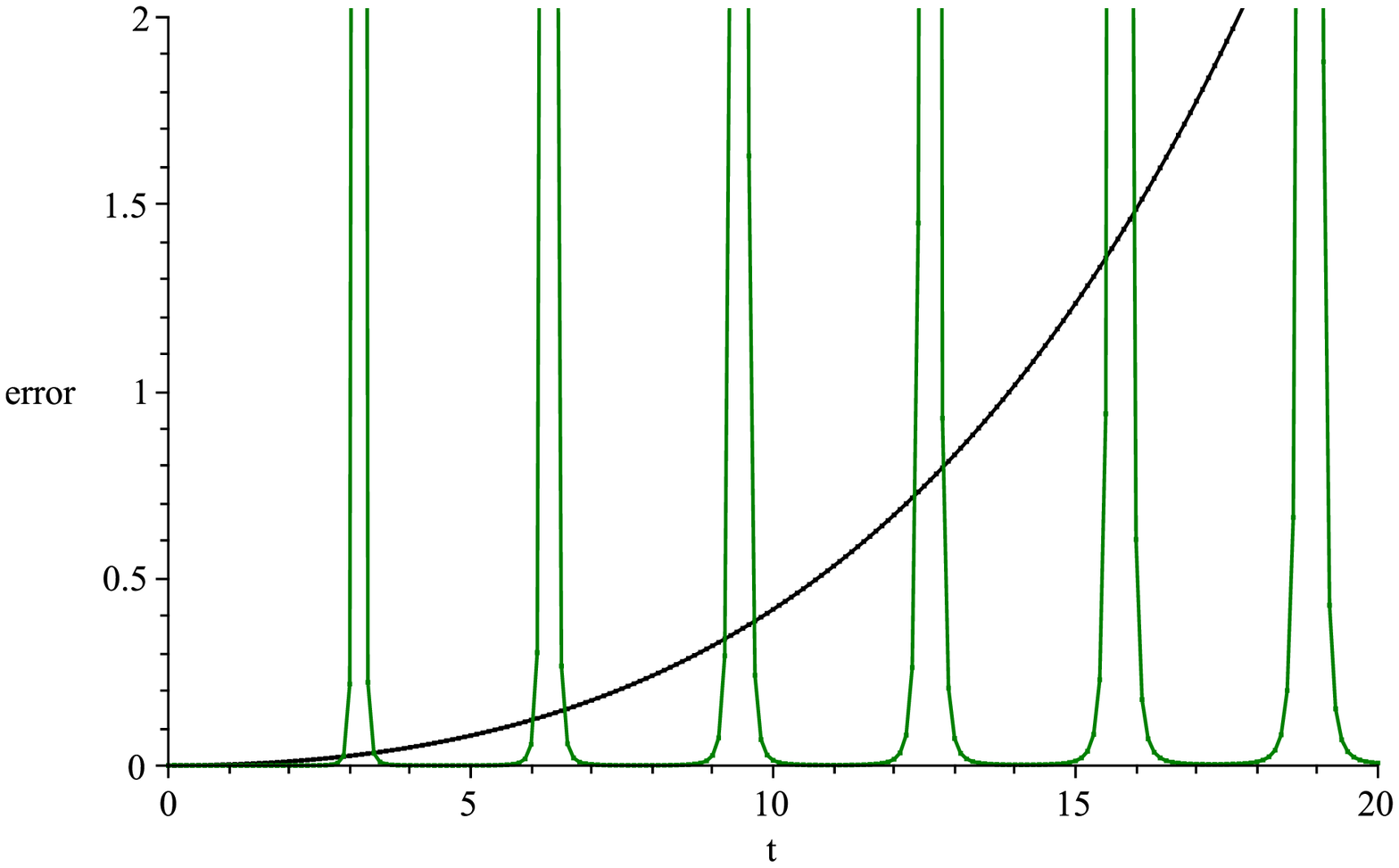}
\label{Euler-error}
\end{figure}

To illustrate the results, Figure \ref{Euler-phase} is the trajectory in the phase space, Figure \ref{Euler-x} is the evolution of $x$ in time, Figure \ref{Euler-p} is the evolution of $p$ in time and Figure \ref{Euler-inv} is the evolution of the invariant $\frac{x}{p}$ in time. In Figure \ref{Euler-error} the error over time in the phase space $\sigma(x,p)$ is given by the black curve while the error over time of the invariant $\sigma(\frac{x}{p})$ is given by the green curve,
\begin{eqnarray}
\sigma(x,p)&=&(x(t)-x(n\Delta))^2+(p(t)-p(n\Delta))^2,\\
\sigma\left(\frac{x}{p}\right)&=&\left(\frac{x(t)}{p(t)}-\frac{x(n\Delta)}{p(n\Delta)}\right)^2
\end{eqnarray}
such that $t=n\Delta$.


We can see in Figure \ref{Euler-error} that the ratio $\frac{x}{p}$ provides a better approximation than the set of variables $x$ and $p$. (Of course, the approximation of the invariant is not good when $p$ approaches zero since it hits a singularity, but around those points, one can use the inverse function $\frac{p}{x}$ to get a better approximation.) In addition, we can see in Figure \ref{Euler-inv} that the blue dots are drifting to the right of the curve. This can be explained by higher-order errors generating an extra time translation.

\pagebreak

In this case, if one corrects the scheme by subtracting the errors $v_2$, $v_3$ and $v_4$ of the Euler scheme, we obtain
\begin{eqnarray}
X=x+\Delta p-\frac{\Delta^2}{2}x-\frac{\Delta^3}{6}p+\frac{\Delta^4}{24}x,\qquad P=p-\Delta x-\frac{\Delta^2}{2}p+\frac{\Delta^3}{6}x+\frac{\Delta^4}{24}p,
\end{eqnarray}
which is equivalent to the Runge--Kutta scheme (RK4), i.e.
\begin{eqnarray}
\frac{X-x}{\Delta}=p-\frac{\Delta}{2}x-\frac{\Delta^2}{6}p+\frac{\Delta^3}{24}x,\qquad \frac{P-p}{\Delta}=-x-\frac{\Delta}{2}p+\frac{\Delta^2}{6}x+\frac{\Delta^3}{24}p.
\end{eqnarray}

\subsubsection{The discrete gradient method}
Let us consider the discrete-gradient method which preserves the Hamiltonian (\ref{Hex2}), i.e. the discrete equations of motion takes the form
\begin{eqnarray}
\frac{X-x}{\Delta}=\frac{1}{2}\frac{P^2-p^2}{P-p},\qquad \frac{P-p}{\Delta}=-\frac{1}{2}\frac{X^2-x^2}{X-x}
\end{eqnarray}
and the iteration scheme is given by
\begin{eqnarray}
X=\frac{4x+4\Delta p-\Delta^2 x}{4+\Delta^2},\qquad P=\frac{4p-4\Delta x-\Delta^2p}{4+\Delta^2}.
\end{eqnarray}
It is interesting to note that for this scheme, the inverse transformation is the same as inverting the sign of $\Delta$. However, it is not a time-linear Lie-group transformation since the combination of two iterations is not the iteration of the sum of the steps, e.g.
\begin{eqnarray}
\psi_{-\Delta}\circ\psi_\Delta=1,\qquad \psi_\Delta\circ\psi_\Delta\circ x\neq F_j(x,p;2\Delta)\quad\psi_\Delta\circ\psi_\Delta\circ p\neq G_j(x,p;2\Delta).
\end{eqnarray}
Applying two iterations with the same step $\Delta$ is equivalent to applying one iteration with the step $\frac{8\Delta}{4-\Delta^2}$.

By direct substitution into equation (\ref{Err}), we obtain
\begin{eqnarray}
\psi_{-\Delta}\circ\frac{\partial}{\partial\Delta}\psi_\Delta=\left(1+\frac{\Delta^2}{4+\Delta^2}\right)\left(p\frac{\partial}{\partial x}-x\frac{\partial}{\partial p}\right).\label{ErrDG}
\end{eqnarray}
and the lower-order terms take the form
\begin{eqnarray}
v_0&=&v_1\;=\;v_2\;=\;0,\nonumber\\
v_3&=&\frac{-1}{2}\left(p\frac{\partial}{\partial x}-x\frac{\partial}{\partial p}\right),\nonumber\\
v_4&=&2\left(x\frac{\partial}{\partial x}+p\frac{\partial}{\partial p}\right),\nonumber\\
v_5&=&\frac{13}{2}\left(p\frac{\partial}{\partial x}-x\frac{\partial}{\partial p}\right),\nonumber\\
&\vdots&\nonumber
\end{eqnarray}
We can see that the error is of order 3 and higher and the first term is proportional to the deformation $\mathfrak{g}$. As we can see, the leading-order error is a translation in time, that is the numerical solution will remain on the trajectory, but it represents a point too far in time. By computing the solution for greater time, we can see that the graphs of the position and momentum in time will slowly drift to the right of the exact curves.

\begin{figure}[h!]
\centering
\caption{}
\includegraphics[scale=0.4]{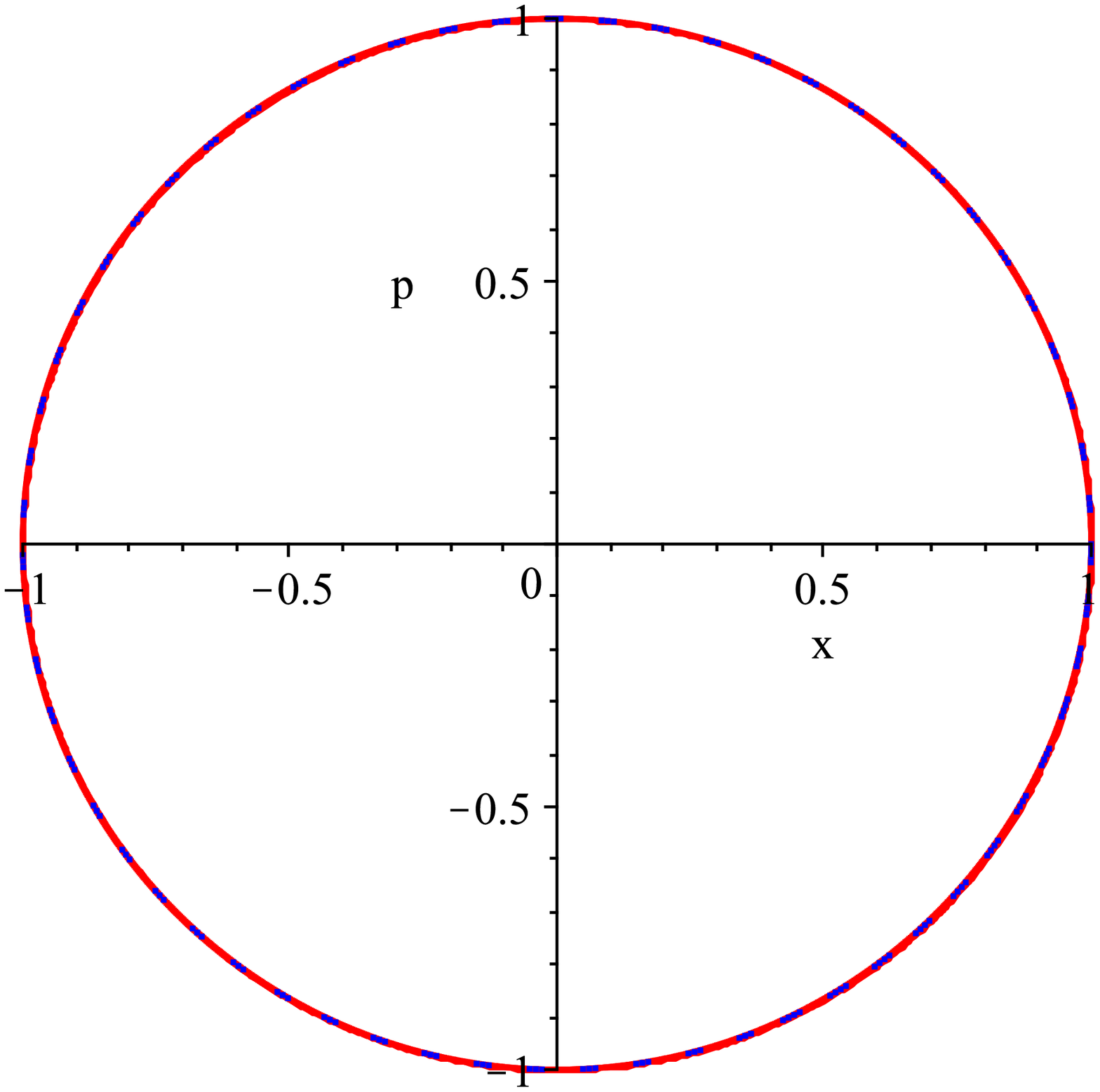}
\label{DisGra-phase}
\end{figure}

\begin{figure}[h!]
\centering
\caption{}
\includegraphics[scale=0.7]{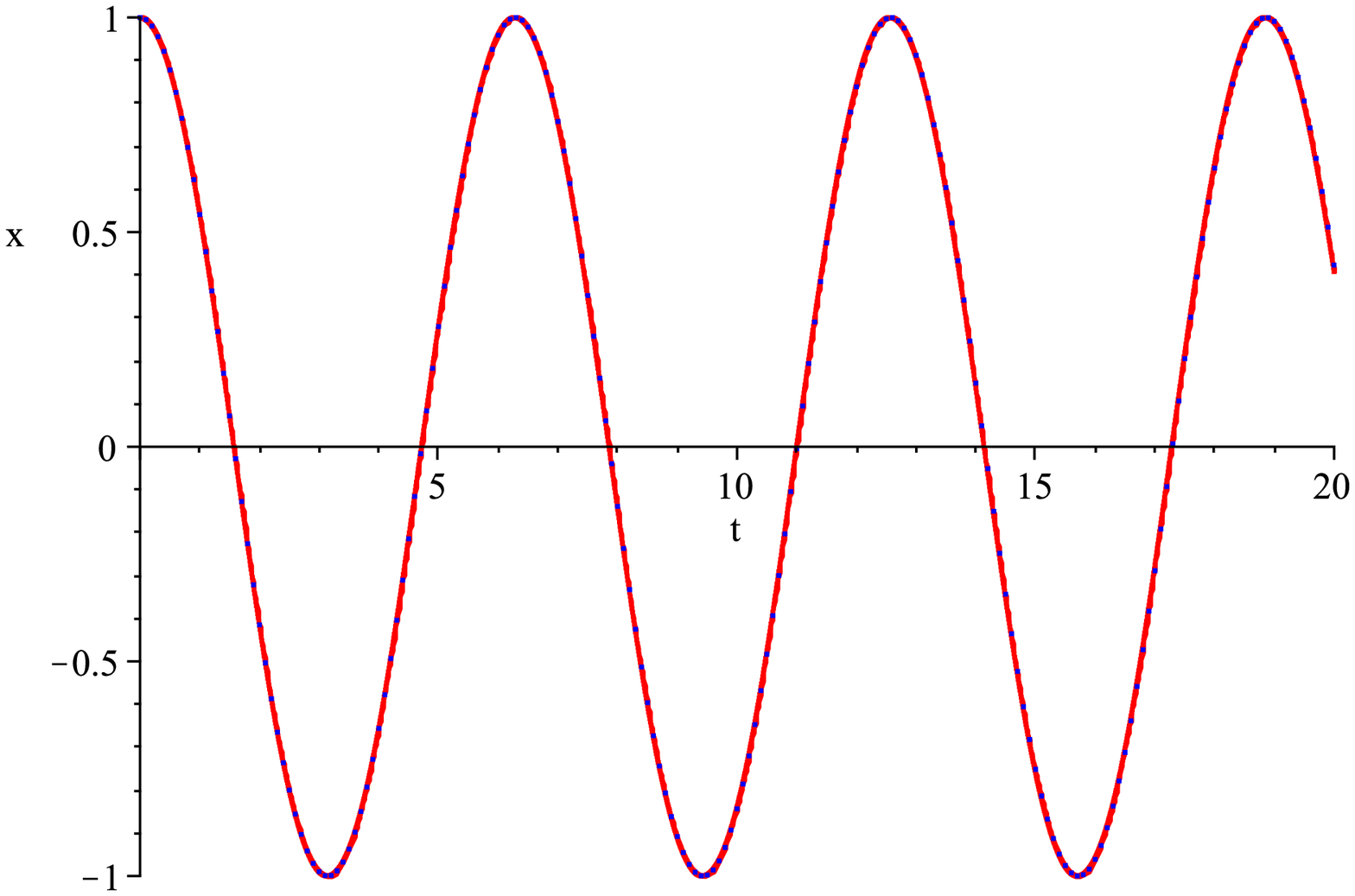}
\label{DisGra-x}
\end{figure}

\begin{figure}[h!]
\centering
\caption{}
\includegraphics[scale=0.7]{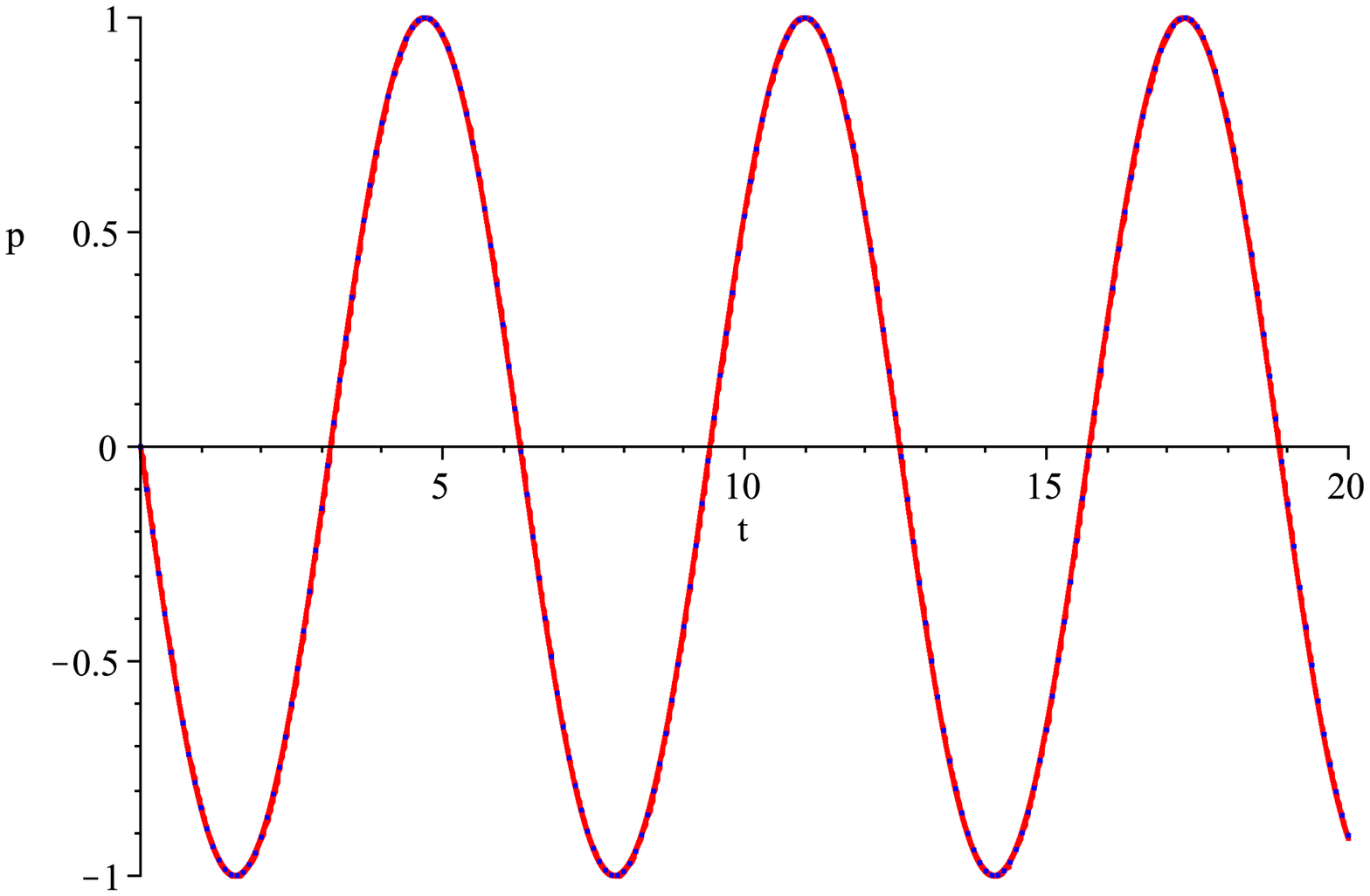}
\label{DisGra-p}
\end{figure}

\begin{figure}[h!]
\centering
\caption{}
\includegraphics[scale=0.7]{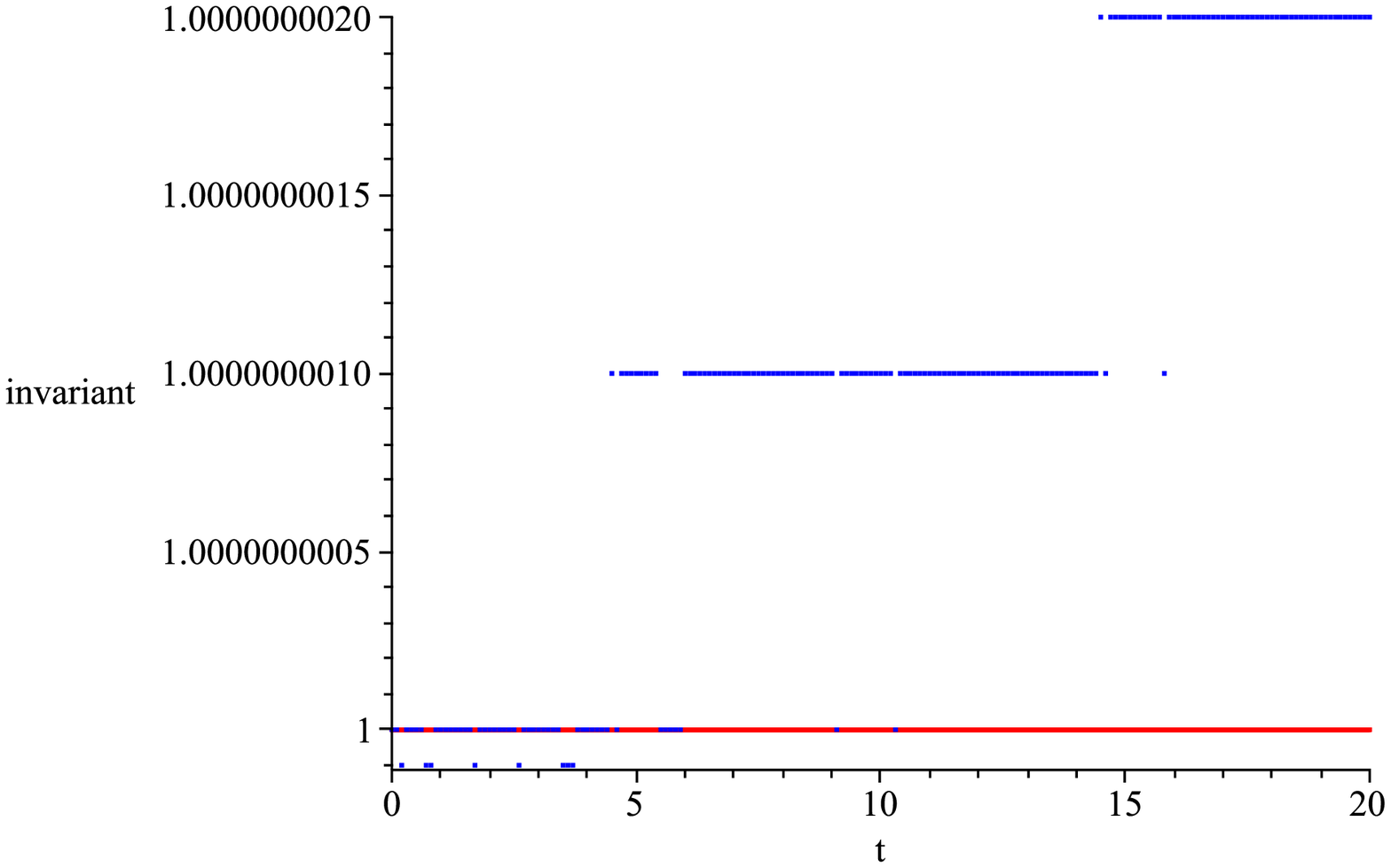}
\label{DisGra-inv}
\end{figure}

\begin{figure}[h!]
\centering
\caption{}
\includegraphics[scale=0.7]{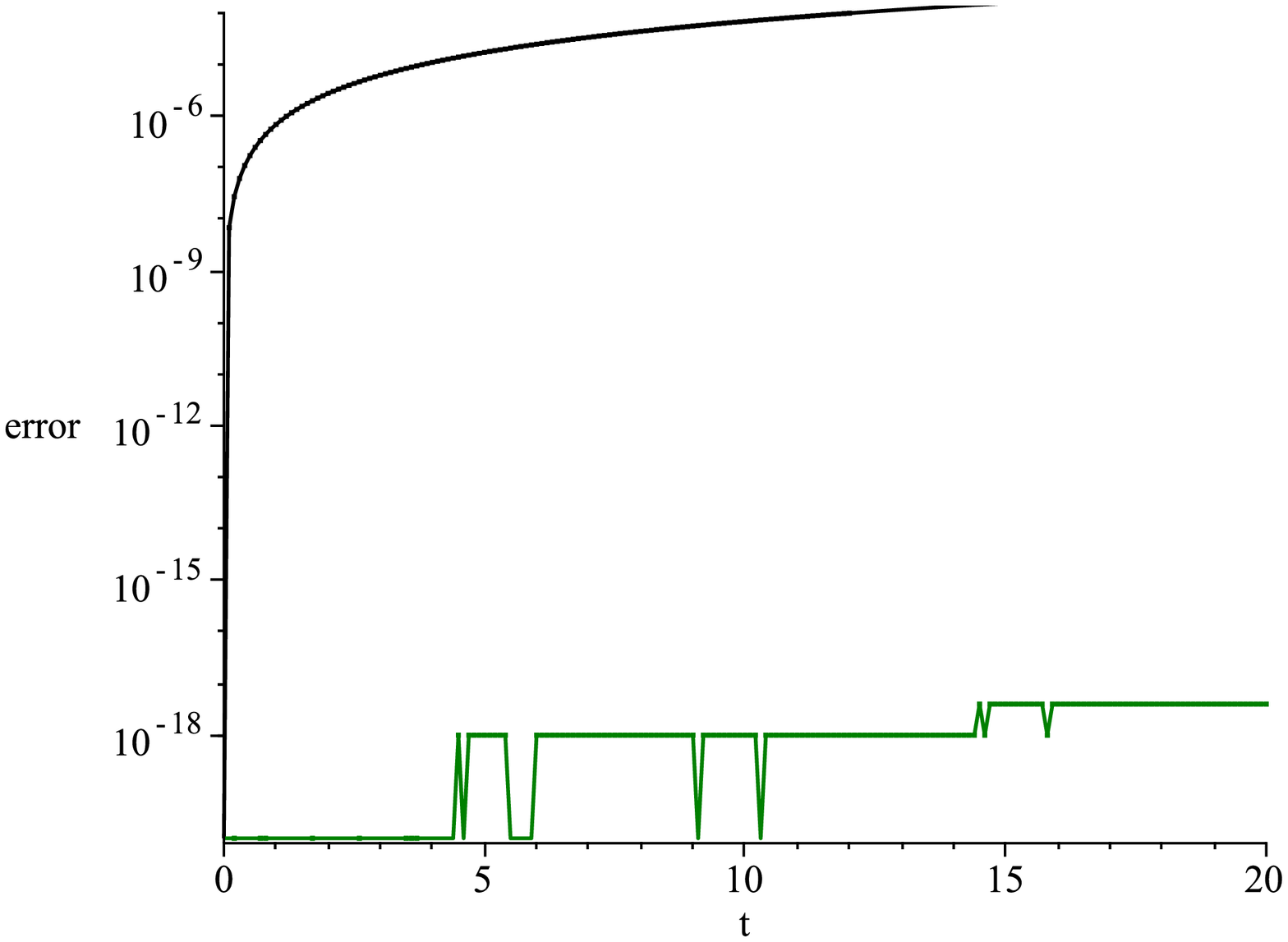}
\label{DisGra-error}
\end{figure}

To illustrate the results, Figure \ref{DisGra-phase} is the trajectory in the phase space, Figure \ref{DisGra-x} is the evolution of $x$ in time and Figure \ref{DisGra-p} is the evolution of $p$ in time and Figure \ref{DisGra-inv} is the evolution of $H$ in time. In Figure \ref{DisGra-error} the error over time in the phase space $\sigma(x,p)$ is given by the black curve while the error over time of the invariant $\sigma(2H)$ is given by the green curve,
\begin{eqnarray}
\sigma(x,p)&=&(x(t)-x(n\Delta))^2+(p(t)-p(n\Delta))^2,\\
\sigma(2H)&=&\left(p(t)^2+x(t)^2-p(n\Delta)^2-x(n\Delta)^2\right)^2
\end{eqnarray}
such that $t=n\Delta$.


\pagebreak

~

\pagebreak

Knowing that the scheme is a non-linear in time group transformation and that the errors solely take the form of extra translation in time, we can express the approximative scheme differently, i.e.
\begin{eqnarray}
\psi_\Delta = \exp\left((\Delta+W(\Delta))\mathfrak{g}\right),
\end{eqnarray}
where $W(\Delta)$ is a function of $\Delta$ representing the error on time. Using (\ref{ErrDG}) and the new definition, we can calculate $W(\Delta)$ to obtain
\begin{eqnarray}
W(\Delta)=\Delta - 2 \arctan\left(\frac{\Delta}{2}\right).
\end{eqnarray}
Hence, by breaking the assumption that $\Delta$ acts as the time $t$, we can correct the scheme by parametrizing the time by a function of $\Delta$. The resulting scheme will be exact, i.e.
\begin{eqnarray}
X=\frac{4x+4\Delta p-\Delta^2 x}{4+\Delta^2},\qquad P=\frac{4p-4\Delta x-\Delta^2p}{4+\Delta^2}, \qquad T=t+ 2 \arctan\left(\frac{\Delta}{2}\right),\nonumber
\end{eqnarray}
where $X$, $P$ and $T$ are the ``advance-time'' coordinates in the augmented phase space.

\section{Conclusions}\label{SecConc}
As a summary, we constructed a formalism allowing us to express solutions of time-independent Hamiltonian systems as deformation of the initial conditions via a Lie-group transformation. We found explicitly the associated Lie algebra, which allows us to compare approximative/numerical schemes with the exact scheme or real solution. This comparison provide information on the amplitude and type of errors.

More precisely, in section \ref{SecForm} we constructed the formalism for time-independent Hamiltonian systems using Lie-group transformations and an evolution generator taking values into the associated Lie algebra. This formalism considers an implicit advance in time, i.e. using the dependent variables, instead of an explicit advance in time, that is transforming solely the time. Under the canonical transformation conditions, the algebraical preservation of all integrals of motion and an equivalent of a time translation, we were able to determine the evolution generator. This vector field generates a Lie-group transformation which is consistent with both continuous and discrete versions of time. The mesh of the discretization is left without constraints. One of the advantages of this method is that we do not require to discretize differential equations and we look for an algebraic substitute. Nothing is needed to be known about the system (except the Hamiltonian and the initial conditions).

In section \ref{SecExact}, we propose a method to construct exact schemes for time-independent Hamiltonian systems that are integrable. These schemes keep the full precision of the continuous case with an arbitrary mesh. However, this method also possesses some problems. Some integrals of motion must be known, which can be a tremendous task by itself. The method can also hit some computational problems when one has to find the action-angle coordinates  and/or the (inverse) canonical transformations, i.e. to give the relation between the original coordinates and the action-angle coordinates.

In section \ref{NonExact}, we investigated non-exact schemes using the formalism of section \ref{SecForm}. This allows us the determine the errors coming from the numerical schemes. It is possible to correct them or look for functions of the coordinates (invariant of the errors) that are more precise. In addition, we showed how the Euler method can be derived from this formalism. We illustrated these considerations via the 1-dimensional harmonic oscillator.  For the Euler method, we were able to correct the scheme to get RK4. We also looked into the discrete gradient method which preserves the Hamiltonian.

This Lie-algebra formalism can be extended in many directions: for one, providing a similar proof of this formalism for autonomous systems of ordinary differential equations. It would be of great use to generalize the formalism to include time-dependent systems. Also, it would be interesting to approach perturbation theory using this method, that is the first-order error is non-zero. It could be also interesting to break the relation between the group parameter and the time, allowing for implicit parametrization of the spacetime. In a more general way, ordinary differential equations are often used as a basis for partial differential equations. It would be interesting to further extend this formalism to partial differential equations. Also, it could have some applications with delay equations. Lastly, a geometric study of the errors of different schemes could be undertaken to investigate the global errors or methods to get exact schemes from non-exact schemes using additional numerical analysis tools.

\section*{Acknowledgements}
SB was partially supported by postdoctoral fellowships provided by the Fonds de Recherche du Qu\'ebec : Nature et Technologie (FRQNT) and the Natural Sciences and Engineering Research Council of Canada (NSERC). SB would like to thank Libor~\v{S}nobl (\v{C}VUT, Czech Republic) and Adel~F.~Antippa (UQTR, Canada) for interesting discussions on the subject of this paper.

\end{document}